\documentstyle[preprint,prb,aps,epsf]{revtex}
\def\q{{\bf{q}}}
\def\k{{\bf{k}}}
\def\Q{{\bf{Q}}}
\def\kl{{\bf{k'}}}
\begin{document}

\draft
\title
{\bf Microscopic Theory of Dipole-Exchange Spin Waves in
Ferromagnetic Films: Linear and Nonlinear Processes}
\bigskip
\author{R. N. Costa Filho$^{1,2,*}$,M. G. Cottam$^2$, and G. A.
Farias$^1$}

\address{$^1$Departamento de F\'\i sica, Universidade Federal do Cear\'a,
Campus do Pici, \\ Centro de Ci\^encias, Caixa Postal 6030,
60455-760 Fortaleza, Cear\'a, Brazil}
\address{$^2$Department of Physics and Astronomy, University of Western Ontario\\ London,
Ontario, Canada N6A 3K7}  \maketitle

\begin{abstract}

The linear and nonlinear processes in ferromagnetic films at low
temperatures ($T\ll T_c$) are studied in a microscopic theory.
Both the long-range magnetic dipole-dipole and the Heisenberg
exchange interactions to nearest and next-nearest neighbors are
included. The results obtained for the linearized spin-wave
spectrum are compared with previous macroscopic theories. For
ultrathin films (or for large wave vectors) the microscopic theory
provides important corrections. The nonlinear dynamics of the spin
waves are studied through a finite-temperature perturbation theory
based on Feynman diagrams. We obtain explicit results for the
energy shift and damping (or reciprocal lifetime) of the
dipole-exchange spin waves due to all possible three-magnon and
four-magnon processes involving combinations of the surface and
quantized bulk spin waves at low temperatures. To investigate
different dipole interaction strengths (relative to the exchange)
numerical results are presented using parameters for Fe, EuO, and
GdCl$_3$.

\end{abstract}


\section{INTRODUCTION}

Ultrathin ferromagnetic films are of great interest due to their
widespread technological applicability.$^1$ This is because these
films can have properties, both static and dynamic, that differ
significantly from bulk samples of the same material. For example,
it is well known theoretically that, in these
quasi-two-dimensional systems, the short-range exchange
interactions alone are not sufficient to establish a
ferromagnetically ordered ground state,$^{2,3}$ and it is
necessary to take into account the anisotropy and long-range
character of the dipolar interactions. Regarding the spin dynamics
in ferromagnetic films, a macroscopic (or continuous medium)
theory for the dipole-dominated regime was given by Damon and
Eshbach$^4$ in terms of magnetostatic modes. They identified a
surface branch to the spectrum, now known as the Damon-Eshbach
(DE) mode. Subsequently there has been extensive work to
generalize the magnetostatic theory to include also the exchange
effects, leading to dipole-exchange theory (for reviews see, e.g.,
Refs. 5-7). In these theories the DE surface mode becomes modified
and various quantization effects become evident for the bulk
modes. In addition, Benson and Mills$^8$ considered a microscopic
formalism for the dipole interactions in a thin ferromagnetic
film, obtaining a rapidly converging series for the dipole sums.
More recently, their approach has been extended to a description
of the dipole-exchange spin waves (SW) in various ferromagnetic
and antiferromagnetic films.$^{9-11}$

All of the studies mentioned above neglected the higher-order
effects that give rise to SW interactions. These occur because the
SW are not exact eigenstates of the magnetic Hamiltonian and they
are the source of several nonlinear effects,$^{6,7}$ such as
parallel pumping of SW, auto-oscillations and the transition to
chaos, soliton formation, and energy renormalization and damping
of the SW modes. Previous studies of the SW interactions {\it in
films} were carried out from a macroscopic point of view,$^{5-7}$
where the dipolar terms were introduced using Maxwell's equations.
Again, this should be appropriate for small wave vectors and for
moderate film thickness.

In infinite ferromagnets, either with or without the inclusion of
the dipolar interactions, the treatment of the interactions
between the bulk SW modes is well established.$^{12}$ The dominant
contributions at $T\ll T_c$ come from three-magnon confluence and
splitting processes and from four-magnon scattering processes. In
the special case of a Heisenberg ferromagnet, where there are only
exchange interactions, the three-magnon process is absent.
However, for a finite ferromagnet, the presence of surfaces leads
to a richer SW spectrum that may consist of one or more localized
surface SW modes as well as the bulk modes (which become quantized
in a film geometry). The occurrence of surface and modified bulk
modes in a finite ferromagnet will give rise to more complicated
schemes or SW interactions, since the scattering processes may now
involve both types of modes.

There has previously been a number of calculations for SW
interactions in semi-infinite {\it Heisenberg} ferromagnets  at
low temperatures $T\ll T_c$ using a Hamiltonian approach. In
particular, results for the damping and the energy renormalization
of the surface SW were obtained in some special cases by Tarasenko
and Kharitonov$^{13}$ and Mazur and Mills.$^{14}$ More detailed
results, which were applicable to both surface and bulk (volume
SWs and induced pinning through surface anisotropy and modified
surface exchange, were derived by Kontos and Cottam$^{15,16}$
using a diagrammatic perturbation method.

In the above-mentioned studies for Heisenberg systems, the
dipole-dipole interactions were ignored, leading to a description
of SW interactions at $T\ll T_c$ only in terms of the
leading-order four-magnon processes. When the dipolar terms are
included there are also three-magnon processes involving the
splitting of SW into two SW modes and the confluence of two SW
modes into a single SW. For example, Rahman and Mills$^{17}$
calculated the renormalized energies for the bulk and
Damon-Eshbach surface SW in the dipolar-exchange regime. For
ferromagnetic films in this regime of long wavelengths (or small
wave vectors), there have been extensive calculations of the
nonlinear SW properties using a macroscopic (or continuous-medium)
type of theory.$^{6,7,17,18}$ A nonlinear {\it microscopic}
theory, based on a Hamiltonian approach, allows us to study the SW
interactions and the renormalization of the discrete SW modes.

The aim of this paper is to provide a {\it microscopic} theory to
calculate linear and nonlinear properties of the localized surface
SW and the quantized bulk SW in ferromagnetic films. This is done
for all wave-vector values and for any film thickness. The theory
includes the effects of the short-range exchange coupling and the
long-range dipole-dipole interactions between spins. The latter
interactions are evaluated in a similar manner to recent work for
ferromagnetic and antiferromagnetic films in the {\it linear} SW
regime.$^{9-11}$ We obtain the dependence of the SW energies (for
each of the discrete branches) on the number of layers of the
film, the variation of the exchange parameters (including nearest
and next-nearest neighbors), the strength of the dipolar
interaction, and the in-plane wave vector. We also develop a
perturbation formalism to study the linear and leading-order
nonlinear SW processes of in ferromagnetic thin films at low
temperature $T\ll T_C$. We use a Hamiltonian formalism to find
formal expressions for the SW interaction terms involving products
of three and four boson operators, representing the three- and
four-magnon processes respectively.  The method basically involves
diagonalizing the non-interacting part of the Hamiltonian that is
bilinear in the boson operators by making a transformation to a
new set of boson operators, and then using it to define
``unperturbed'' Green functions. The perturbation terms, involving
either three or four boson operators, are each transformed to the
new boson operators and used to define interaction vertices within
a Feynman diagram formalism.

This paper is arranged as follows. Section II describes the
Hamiltonian with exchange, Zeeman, and dipolar terms, as well as
the assumed geometry of the ferromagnetic film. The method
basically involves generalizing some previous calculations for
nonlinear SW processes in semi-infinite Heisenberg
ferromagnets$^{15}$ to include the dipolar interactions and the
effect of finite film thickness. The transformations of the dipole
sums that enter into our Hamiltonian were calculated according to
the method described in Ref. 8. Results for the linear
(noninteracting) SW spectrum of the film are derived in Sec. III,
where we include numerical applications to different materials. In
Sec. IV we analyze the higher order terms in the Hamiltonian that
lead to SW interactions. At low temperatures $T \ll T_c$, it is
shown that the dominant interaction processes involve three- and
four- magnon scattering, by analogy with infinite
ferromagnets.$^{12}$. In Sec. V we introduce the Green function
formalism, associated with the diagonalized Hamiltonian $H^{(2)}$
of the system, and employ it to obtain general expressions for the
energy shift and damping of the magnons in a ferromagnetic film in
terms of a SW self-energy contribution. A diagrammatic
perturbation technique is then employed in Sec. VI to calculate
the SW self-energy due to the three- and four-magnon  processes.
In Sec. VII we discuss the numerical results applied to three
different materials: Fe, where the ratio of dipolar to exchange
strengths is very small; GdCl$_3$, where the dipolar terms are
relatively strong; and EuO, which represents an intermediate case.
The overall conclusions are in Sec. VIII, while some of the
mathematical expressions are quoted in the Appendices.


\section{HAMILTONIAN AND FILM GEOMETRY}

We consider a ferromagnetic film with $N$ atomic layers of spins
(with quantum number $S$) arranged on a simple cubic lattice with
lattice constant $a$. The external magnetic field $H_0$ and the
static magnetization $M$ are assumed to be parallel to $z$
direction, the surface of the film is in the $x-z$ plane, while
the $y$-axis is perpendicular to the surface. We can generalize to
the other crystal structures, surface orientations, and
magnetization directions in a straightforward manner.

The system will be represented by the Hamiltonian

\begin{equation}
H=-\frac 12\sum_{ij}J_{ij}{\bf S}_i\cdot{\bf
S}_j-g\mu_BH_0\sum_iS_i^z+\frac
12(g\mu_B)^2\sum_{ij}D_{ij}^{\alpha\beta}S_i^{\alpha}S_j^{\beta}
\end{equation}
where ${\bf S}_i$ is the spin at site $i$, and $J_{ij}$ is the
exchange coupling between sites labeled $i$ and $j$. It is assumed
that the exchange coupling is $J_1$ and $J_2$ for nearest and
next-nearest neighbors respectively, and zero otherwise. The
second term of Eq. (1) describes the Zeeman interaction of the
spins with an externally applied field $H_0$. The dipole-dipole
interaction between spins in different lattice sites is
represented by the last term in Eq. (1), where $\alpha$ and
$\beta$ denote components $x$, $y$, or $z$, and

\begin{equation}
D_{ij}^{\alpha\beta}=\left\{\frac {|{\bf
r}_{ij}|^2\delta_{\alpha\beta}-3r_{ij}^{\alpha}r_{ij}^{\beta}}
{|{\bf r}_{ij}|^5} \right\}
\end{equation}
where ${\bf r}_{ij}={\bf r}_j-{\bf r}_i$, and the case $i=j$ is
excluded from the sums in Eq. (1). Here we consider, for
simplicity, that the exchange constants do not change near the
surface of the film, and surface anisotropy fields are also
neglected. However, we can easily include these changes in our
calculations.

The Hamiltonian may be written in terms of boson operators using
the Holstein-Primakoff transformation,$^{12}$ and can be expanded
as $H=H^{(2)}+H^{(3)}+H^{(4)}+\ldots$ (apart from a constant),
where $H^{(m)}$ denotes the term involving a product of $m$ boson
operators. The non-interacting (linear) SW modes are described by
the quadratic term $H^{(2)}$  that has the following form

\begin{equation}
H^{(2)}=\sum_{\q nn'}\left\{ A_{nn'}^{(2)}(\q)a_{\q n}^{\dagger }
a_{\q n'}+B_{nn'}^{(2)}(\q)[ a_{\q n}a_{-\q n'}+a_{\q
n}^{\dagger}a_{-\q n'}^{\dagger}]\right\}.
\end{equation}
Here we are using a representation of the boson operators
$a^{\dagger}$ and $a$ in terms of a 2D wave vector $\q = (q_x,
q_z)$ parallel to the film surfaces and indices $n$ and $n' (= 1,
2,\ldots, N)$ that label the atomic layers parallel to the
surface. The amplitude factors are

\begin{eqnarray}
A_{nn'}^{(2)}(\q) &=&\left\{ g\mu_BH_0
+S\left[u_n(0)-u_n(\q)+v_{n,n+1}(0)+v_{n,n-1}(0)-( g\mu_B)^2
\sum_{n''}D_{nn''}^{zz}(0)\right]\right\}\delta _{nn'} \nonumber
\\ &&-S[v_{n,n+1}(\q)\delta _{n',n+1} +v_{n,n-1}(\q)\delta
_{n',n-1}+(g\mu _B)^2D_{nn'}^{zz}(\q)] ,
\end{eqnarray}

\begin{equation}
B_{nn'}^{(2)}(q)=\frac 14 S(g\mu_B)^2
\left[D_{nn'}^{xx}(\q)-D_{nn'}^{yy}(\q)-
2iD_{nn'}^{xy}(\q)\right].
\end{equation}
Here we have introduced

\begin{equation}
u_n (\q) = 2J_1\sigma_1(\q) + 4J_2\sigma_2(\q)
\end{equation}
and

\begin{equation}
v_{n,n\pm 1}(\q) = J_1 + 2J_2\sigma_1(\q)
\end{equation}
as intra-layer and adjacent-layer Fourier transforms of the
exchange interactions, with $\sigma_1(\q) = [\cos(q_xa) +
\cos(q_za)]$ and $\sigma_2(\q) = \cos(q_xa)\cos(q_za)$.  The
quantities $D_{nn'}^{\alpha\beta}$ are analogous Fourier
transforms with respect to $\q$ of the dipole interactions
defined in Eq. (2). The expressions for these terms are given in
Appendix A.

In the next section we discuss the linearized SW spectrum, which is
obtained using $H^{(2)}$, before proceeding in Sec. IV to a
description of the nonlinearities produced by the higher-order
terms.


\section{LINEARIZED SW SPECTRUM}

In the linear approximation the SW spectrum is obtained by
applying the standard equation of motion $i\hbar dA/dt =  [A,H]$
(for any operator $A$) to the boson operators $a_{\q n}^{\dagger}$
and $a_{\q n}$ in layer $n$, with the Hamiltonian $H$ replaced by
$H^{(2)}$. Taking $\hbar =1$ and assuming a time dependence for
the modes like $\exp (-i\omega t)$, we obtain the sets of
equations

\begin{equation}
\omega a_{\q n}^{\dagger}=\sum_{n'}\left\{A_{nn'}^{(2)}(\q)a_{\q
n}^{\dagger}+[B_{nn'}^{(2)}(\q)+B_{n'n}^{(2)}(-\q)]a_{-\q
n'}\right\},
\end{equation}

\begin{equation}
-\omega a_{-\q n'}=\sum_{n'}\left\{[B_{nn'}^{(2)}(-\q)+
B_{n'n}^{(2)}(\q)] a_{\q n}^{\dagger}+A_{nn'}^{(2)}(-\q)a_{-\q
n'}\right\}.
\end{equation}

There are $2N$ coupled equations altogether, where $N$ is the
number of layers of the film. Using the symmetry properties
$A_{nn'}^{(2)}(\q)=A_{nn'}^{(2)}(-\q)$ and
$B_{n'n}^{(2)}(-\q)=B_{nn'}^{(2)}(\q)$, which arise from the
properties of the dipole sums (see Appendix A) and the property
$J_{ij}=J_{ji}$, the condition for there to be non-trivial
solutions of Eqs. (8) and (9) can be expressed as
\begin{equation}
\det\left[
\begin{array}{cc}
{\bf A}^{(2)}(\q)-\omega {\bf I}_N & 2\tilde{{\bf B}}^{(2)}(\q)\\ 2{\bf
B}^{(2)}(\q)  & {\bf A}^{(2)}(\q)+\omega {\bf I}_N
\end{array}
\right] =0.
\end{equation}
Here ${\bf A}^{(2)}(\q)$ and ${\bf B}^{(2)}(\q)$ are $N\times N$
matrices with elements defined in Eqs. (4) and (5) respectively,
and ${\bf I}_N$ is the $N\times N$ unit matrix. The discrete SW
frequencies in the film correspond to the $N$ solutions for
positive $\omega$ of Eq. (10). They are degenerate in magnitude
with the negative-frequency solutions.

Numerical applications of the above theory have been made using
parameters appropriate to thin films of the ferromagnets GdCl$_3$
($T_c \approx 2.2$ K) and EuO ($T_c \approx 69$ K). These are
chosen as representing different strengths of the dipole
interactions relative to the exchange interactions. They
correspond, in fact, to increasing exchange strengths, as can be
inferred from the increasing $T_c$ values. In general, the dipolar
interactions will have their most significant effect on the
long-wavelength (small $\q$) SW properties in each material, while
the exchange effects dominate at shorter wavelengths. We note that
thin-film samples of EuO have been employed in SW experiments,
e.g., using Brillouin light scattering and microwave resonance
techniques.$^{19,20}$ Consequently, their magnetic parameters are
fairly well known. In the case of GdCl$_3$ the approximate
parameters can be deduced from a study of the SW spectrum in bulk
samples (e.g., see Ref. 21). The ratio of $4\pi M$ (which
characterizes the dipolar strength) to the bulk exchange field
$H_{Ex}$ is approximately 1.5 for GdCl$_3$ and 0.063 for EuO.

It is well known that, in Heisenberg ferromagnetic films,
exchange-dependent surface SW modes may occur$^{5,7}$ that have
properties quite different from the Damon-Eshbach (DE) surface
mode$^{4,5}$ of the magnetostatic theory. They occur, for example,
if the exchange coupling between spins in the surface layers is
modified or if effects of next-nearest neighbor exchange
interactions are included. Here we study such exchange surface
modes (as well as the DE type of surface modes) by including the
next-neighbor exchange interactions $J_2$ in our numerical
calculations for ultrathin films.

We first show calculations for a 16-layer EuO film, taking $4\pi M
= 2.4$ T, $H_{Ex} = 38$ T, and $H_0 = 0.36$ T. Note that the bulk
exchange field is given by $g\mu_BH_{Ex} = 6S(J_1+2J_2)$ for the
model simple-cubic structure. Here $J_1$ and $J_2$ are effective
exchange parameters because EuO normally has a fcc structure. In
Fig. 1 plots are shown for the frequency in GHz (converted using
$\gamma = 28$ GHz/T) of the lowest few discrete SW branches versus
the ratio $J_2/J_1$ and versus $q_xa/\pi$ (which ranges from 0 at
the Brillouin zone center to 1 at the zone boundary) for small
$q_x$. We have taken the case of $q_z = 0$ (the Voigt geometry),
for which it is known that there is a Damon-Eshbach (DE) surface
mode$^4$ in the magnetostatic continuum limit where exchange is
neglected. In Fig. 1(a) the dependence of the two lowest branches
of the SW dispersion relation on $J_2/J_1$ is shown for two
different values of $q_xa/\pi$. For $q_x\approx 0$ we note that
the first branch has almost no dependence on $J_2/J_1$. That is
expected since this lowest branch is essentially the DE mode. The
second branch has a smooth dependence on $J_2$. For $q_x=0.01$
both modes increase their energy as $J_2/J_1$ increases. These
effects are more noticeable in Fig. 1(b), where a plot of the
frequency against the wave vector is shown for three different
values of $J_2/J_1$. The purely magnetostatic DE mode would tend
to the approximate limiting value of 43 GHz with increasing $q_x$.
While it can be seen in both cases that there is a relatively flat
(dispersionless) branch near this frequency, the upward curvature
apparent at larger $q_x$ is due to the exchange terms. In the
region near 35 GHz for small $q_xa/\pi$ where two branches (the
lowest bulk mode and the analog of the DE mode) come close
together, but do not cross. This phenomenon of a near
``crossover'', with mode repulsion, signifies a strong mixing (or
hybridization) of two modes.

In Fig. 2 we show the behavior of the lowest four SW branches for
very small wave vectors in a GdCl$_3$ film with $N=16$,
considering $J_2/J_1=0$ (dashed lines) and $J_2/J_1=0.25$ (solid
lines). This shows the effects of the next-nearest neighbor
exchange. For this material, we use the parameters $4\pi M = 0.82$
T, $H_{Ex} = 0.54$ T, and $H_0= 0.36$ T. It can be seen that a
purely magnetostatic DE mode starts at $q_x=0$ at the frequency
$18.25$ GHz in both cases. By contrast with the same film
thickness for EuO in Fig. 1, we can see that the DE mode in the
GdCl$_3$ case affects several of the bulk modes of the spectrum,
which is a consequence of the larger $4\pi M / H_{Ex}$ in this
material compared with EuO.

In the numerical examples shown above, the limiting frequency of
the DE mode (as $q_x \rightarrow 0$) is very close to the value
$[H_0(H_0+4\pi M)]^{1/2}$ predicted by the macroscopic
magnetostatic theory$^4$ in the Voigt geometry. On the other hand,
the magnetostatic theory predicts that the DE branch becomes flat,
tending to a limiting value of $H_0+2\pi M$ when $q_x$ is large
compared with the reciprocal of the film thickness. This behavior
is followed approximately in our theory, but there is an
additional upward curvature to all SW branches due to exchange.
This is in addition to the hybridization effects and the
microscopic corrections to continuum dipole-exchange theories.


\section{HIGHER ORDERS: SW INTERACTIONS}

The higher order terms in the expansion of the Hamiltonian, i.e.,
the terms $H^{(m)}$ with $m \ge 3$, correspond to the nonlinear
aspects of the SW dynamics including interactions between the SW
modes. As in the case of bulk (effectively infinite)
ferromagnets,$^{12}$ the leading-order effects for $T\ll T_C$ come
from the terms that have three and four boson operators ($H^{(3)}$
and $H^{(4)}$).

The term $H^{(3)}$ depends only on the dipolar interactions and has the form:

\begin{equation}
H^{(3)}=\frac 12\sum_{\k\q nn'}A_{nn'}^{(3)}(\k) \left[ a_{\k
n'}^{\dagger }a_{\q-\k n}^{\dagger }a_{\q n}+a_{\q n}^{\dagger
}a_{\q-\k n}a_{\k n'}\right],
\end{equation}
with the amplitude term given by

\begin{equation}
A_{nn'}^{(3)}(\k)=\sqrt{2S}g^2\mu_B^2\left[
D_{nn'}^{xz}(\k)-iD_{nn'}^{yz}(\k) \right].
\end{equation}
This represents the splitting and confluence processes inherent in
the three-magnon processes.

By contrast, the four-magnon interaction $H^{(4)}$ has parts
involving both dipolar and exchange terms:

\begin{eqnarray}
H^{(4)} &=&\frac 12\sum_{\k\k'\q nn'}\left\{
A_{nn'}^{(4)}(\k)a_{\k'n}^{\dagger } a_{\q-\k'n}^{\dagger
}a_{\q-\k n}a_{\k n'}+A_{nn'}^{(4)}(-\k') a_{\k'n'}^{\dagger}
a_{\q-\k'n}^{\dagger} a_{\q-\k n}a_{\k
n}\right.\nonumber\\&&\left. -2B_{nn'}^{(4)}(\k-\k')
a_{\k'n'}^{\dagger}a_{\q-\k'n}^{\dagger } a_{\q-\k n}a_{\k
n'}+C_{nn'}^{(4)}(\k)[ a_{\q n}^{\dagger}
a_{\q-\k-\k'n}a_{\k'n}a_{\k n'}+h.c]\right\}.
\end{eqnarray}
In this case the amplitude coefficients are defined by

\begin{equation}
A_{nn'}^{(4)}(\q)=\frac 12\left[u_n(\q)\delta_{nn'}+v_{n,n+1}(\q)
\delta_{n',n+1}+v_{n',n-1}(\q)\delta _{n',n+1}+\frac 12(g\mu _B)^2
D_{nn'}^{zz}(\q)\right] ,
\end{equation}

\begin{equation}
B_{nn'}^{(4)}(\q) =A_{nn'}^{(4)}(\q)-\frac 34(g\mu_B)^2
D_{nn'}^{zz}(\q),
\end{equation}

\begin{equation}
C_{nn'}^{(4)}(\q) = -(1/S)B_{nn'}^{(2)}(\q).
\end{equation}
This interaction represents the scattering of a pair of magnons
into another pair of magnons.

We note that each term in Eqs. (11) and (13) conserve the 2D
in-plane wave vector, since there is translational symmetry in the
$x-z$ plane.  The dependence in the $y$-direction is taken into
account the the summations over layer indices $n$ and $n'$. The
above $H^{(3)}$ and $H^{(4)}$ interactions are the analogs (in a
microscopic theory) of the expressions used in macroscopic
theories to describe a wide range on nonlinear effects,$^{5,6}$ as
mentioned earlier. For example, they describe processes whereby
energy from a given SW (with, say, wave vector $\q$ and from the
discrete branch $\nu$) can be transferred to other SW modes of the
film. Thus they can be used to calculate the damping (or
reciprocal lifetime) of any given SW, as well as the shift (or
renormalization) in its energy. This can be accomplished using a
diagrammatic perturbation method with Green functions.

\section{THE GREEN FUNCTION FORMALISM}

We establish a Green function formalism for the film geometry with
$N$ atomic layers parallel to the surfaces.  The method is
analogous to that used in previous diagrammatic perturbation
calculations for the effects of SW interactions in semi-infinite
Heisenberg ferromagnets,$^{15,16}$ but generalized here to include
the dipolar terms and the quantization of the modes due to finite
film thickness.

The starting point to the diagrammatic formalism is the expansion
of the Hamiltonian into terms containing products of two, three,
and four boson operators. The relevant expressions for these terms
are given in Eqs. (3), (11), and (13). The bilinear term $H^{(2)}$,
which we shall treat as the non-interacting part of the
Hamiltonian, can first be rewritten as

\begin{equation}
H^{\left( 2\right) }=-\frac 12\sum_{\q}\text{Tr}\left[ {\bf
A}^{\left( 2\right) }\left( \q\right) \right] +\frac 12 \sum_{\q}
\widetilde{{\cal A}}_{\q}^{\dagger }{\bf \ } {\cal X}\left(
\q\right) {\cal A}_{\q},
\end{equation}
where

\begin{equation}
{\cal X}\left( \q\right) =\left(
\begin{array}{ll}
\tilde{{\bf A}}^{\left( 2\right) }\left( \q\right) & 2 \tilde{{\bf
C}}^{\left( 2\right) }\left( \q\right) \\ 2 \tilde{{\bf
B}}^{\left( 2\right) }\left( -\q\right) & {\bf A}^{\left( 2\right)
}\left( -\q\right)
\end{array}
\right).
\end{equation}
Here we have defined operators $\cal{A}_{\q}^{\dagger}$ and
$\cal{A}_{\q}$ that create and destroy wave vector $\q$ by

\begin{equation}
{\cal A}_{\q}^{\dagger }=\left(
\begin{array}{l}
{\bf a}_{\q}^{\dagger } \\ {\bf a}_{-\q}
\end{array}
\right) \qquad ,\qquad {\cal A}_\q=\left(
\begin{array}{l}
{\bf a}_\q \\ {\bf a}_{-\q}^{\dagger }
\end{array}
\right) ,
\end{equation}
respectively, where $a_{\q}^{\dagger}$ and $a_{\q}$ are
$N$-component column matrices whose $n$th elements are $a_{\q,
n}^{\dagger}$ and $a_{\q,n}$. The tilde denotes a matrix
transpose, and ${\bf A}^{(2)}(\q)$ and ${\bf B}^{(2)}(\q)$ are the
$N\times N$ matrices with elements given by $A_{nn'}^{(2)}(\q)$
and $B_{nn'}^{(2)}(\q)$ defined in Eqs. (4) and (5).

The first term in Eq. (17) is just a constant, while the second
term provides the non-interacting (linear) SW spectrum. In fact,
the eigenvalues of $\chi(\q)$ are the discrete SW energies
$E_{\q,\nu}$ where $\nu = 1,2,\dots,N$ is a branch label (each
eigenvalue will occur $twice$ in $\chi(\q)$). Results for
$E_{\q,\nu}$ have already been obtained using the equivalent
determinantal condition in Eq. (10). We now diagonalize $H^{(2)}$
by making a linear transformation to a new set of boson operators
$\alpha_{\q}^{\dagger}$ and $\alpha_{\q}$, which satisfy the usual
commutation relations and are defined by

\begin{eqnarray}
a_{\q,n} &=&\sum_lS_{nl}\left( \q\right) \alpha _{\q
,l}+T_{nl}^{*}\left( \q\right) \alpha _{-\q ,l}^{\dagger },
\\ a_{\q ,n}^{\dagger } &=&\sum_lS_{nl}^{*}\left( \q\right) \alpha
_{\q ,l}^{\dagger }+T_{nl}\left( \q\right) \alpha _{-\q ,l}.
\end{eqnarray}
The $S_{nl}$ and $T_{nl}$ coefficients can be found by a
straightforward generalization of a procedure due to White et
al,$^{22}$ and the result is as follows. Suppose we rewrite the
above transformation in matrix form as
$\tilde{\cal{A}}_{\q}^{\dagger}=
S_{\q}^{\ast}\Lambda_{\q}^{\dagger}$ and
${\cal{A}}_{\q}=S_{\q}\Lambda_{\q}$, where
$\Lambda_{\q}^{\dagger}$ and $\Lambda_{\q}$ are defined similarly
to $\cal{A}_{\q}^{\dagger}$ and $\cal{A}_{\q}$ but in terms of the
new operators.  The $j$th column (denoted by $S_{j,\q}$) of the
$2N\times 2N$ transformation matrix $S_{\q}$ is then found by
solving for the eigenvector in

\begin{equation}
\left(
\begin{array}{cc}
\tilde{{\bf A}}^{\left( 2\right) }\left( \q\right) & 2\tilde{{\bf C}}%
^{\left( 2\right) }\left( \q\right) \\ -2\tilde{{\bf B}}^{\left(
2\right) }\left( -\q\right) & -{\bf A}^{\left( 2\right) }\left(
-\q\right)
\end{array}
\right) {\bf S}_{j,\q}=\pm \epsilon _j\left( \q\right) {\bf
S}_{j,\q},
\end{equation}
where the $+$ sign is taken for $j \in {1,\dots,N}$ and the $-$
sign for $j \in {N+1,\dots,2N}$. The $H^{(2)}$ part of the
Hamiltonian then assumes the simple form (apart from a constant
term):

\begin{equation}
H^{(2)}=\sum_{\q\nu}E_{\q,\nu}(1+2\alpha_{\q\nu}^{\dagger}\alpha_{\q\nu}).
\end{equation}

To establish the diagram technique we now introduce a $N\times N$
matrix ${\bf G}(\q,i\omega _m)$ with matrix elements defined as
the causal Green functions involving the transformed boson
operators:

\begin{equation}
G_{\nu \nu'}(\q,i\omega _m) =\left\langle \left\langle \alpha
_{\q\nu };\alpha _{\q\nu'}^{\dagger }\right\rangle
\right\rangle_{i\omega _m}.
\end{equation}
in a conventional notation.$^{13}$ Here $i\omega _m \equiv 2\pi
mi/\beta$ is an imaginary boson frequency, where $\beta = 1/k_BT$
and $m$ takes all integer values from $-\infty$ to $+\infty$. When
these Green functions are evaluated using just $H^{(2)}$ in Eq.
(23) for the Hamiltonian the resulting non-interacting Green
functions are diagonal in the labels $\nu$ and $\nu'$ and are
given by

\begin{equation}
G_{\nu\nu'}^0(\q,i\omega _m) = G_\nu ^0(\q,i\omega _m)\delta _{\nu
,\nu'} = \left(-\frac 1\beta\right)\frac 1 {i\omega_m-E_{\q,\nu }}
\delta_{\nu ,\nu'}.
\end{equation}

On making an analytic continuation $i\omega _m \rightarrow \omega
+ i0^+$ to real frequency $\omega$, it is seen that there is a
simple pole at the values $E_{\q,\nu }$. These correspond to the
non-interacting excitations, comprising the discrete bulk and
surface SW modes, which we have already been in the linear case.
In the higher orders of perturbation, i.e. when  $H^{(3)}$ and
$H^{(4)}$ are included, the ``interacting'' (or renormalized)
Green functions will have modified poles, corresponding to a
renormalization in energy by an amount $\Delta E_{\q,\nu }$ and a
damping $\Gamma_{\q,\nu }$. Also, in general, the interacting
Green functions will have off-diagonal terms in the $\nu$ and
$\nu'$ labels. As in standard diagrammatic formulations for
interacting boson systems, the interacting and non-interacting
Green functions are related to one another by means of the Dyson
development operator.$^{23}$ This quantity can be expanded as an
infinite series in powers of the higher-order Hamiltonian parts.
The terms in the expansion can conveniently be evaluated using a
diagrammatic representation with proper self-energy terms
$\Sigma_{\nu\nu'}(\q,i\omega _m)$ introduced to account for the
interactions. The choice of the self-energy terms within the
perturbation expansion depends on the explicit form of $H^{(3)}$
and $H^{(4)}$ and will be discussed in the following sections.

First, the non-interacting Green function
$G_{\nu\nu'}^0(\q,i\omega _m)$ will be drawn as a solid directed
line as shown in Fig. 3(a). There are then four possible types of
proper self-energy terms, as defined in Fig. 3(b), depending on
whether the $G^0$ lines are entering or leaving. All of these may
be nonzero when there are magnetic dipole-dipole contributions to
the Hamiltonian. The renormalized Green functions are given by a
generalized Dyson series, which is represented diagrammatically in
Fig. 4. This takes the algebraic form

\begin{eqnarray}
G_{\nu \nu'}(\q,i\omega _m) &=&
G_{\nu}^0(\q,i\omega_m)\delta_{\nu\nu'} + G_{\nu}^0(\q,i\omega
_m)\Sigma_{\nu\nu'}^{+-}(\q,i\omega_m)G_{\nu'}^0 (\q,i\omega
_m)\nonumber\\ &&+ \sum_{\nu''}G_{\nu}^0(\q,i\omega
_m)\Sigma_{\nu\nu''}^{+-}(\q,i\omega_m) G_{\nu''}^0(\q,i\omega
_m)\Sigma_{\nu''\nu'}^{+-}(\q,i\omega_m) G_{\nu'}^0(\q,i\omega
_m)\nonumber\\ &&+ \sum_{\nu''}G_{\nu}^0(\q,i\omega
_m)\Sigma_{\nu\nu''}^{++}(\q,i\omega_m) G_{\nu''}^0(\q,i\omega
_m)\Sigma_{\nu''\nu'}^{--}(\q,i\omega_m) G_{\nu'}^0(\q,i\omega
_m)+...
\end{eqnarray}
We note that the only term linear in the self-energy parts is the
one involving $\Sigma^{+-}$. The terms with $\Sigma^{++}$ and
$\Sigma^{--}$ occur in diagrams quadratic (and higher) in the
self-energy parts, and $\Sigma^{-+}$ occurs only in diagrams that
are cubic and higher. If we ignore, for the moment, the effects of
$\Sigma^{++}$, $\Sigma^{--}$, and $\Sigma^{-+}$, the infinite
Dyson equation series for $G_{\nu \nu'}(\q,i\omega _m)$ is
summable, and the formal solution can be written in a matrix form
as

\begin{equation}
{\bf G}(\q,i\omega _m)={\bf G}^0(\q,i\omega _m)+{\bf
G}^0(\q,i\omega_m){\bf \Sigma }^{+-}(\q,i\omega_m){\bf
G}(\q,i\omega_m) .
\end{equation}
The solution of this matrix equation is

\begin{equation}
{\bf G}(\q,i\omega _m) = \left\{ \left[ {\bf G}^0(\q,i\omega
_m)\right] ^{-1}- {\bf\Sigma }^{+-}(\q,i\omega _m)\right\}^{-1}.
\end{equation}

By solving for the complex poles of this interacting Green
function we may deduce the renormalized SW energies $\Delta
E_{\q,\nu }$ and their damping $\Gamma_{\q,\nu }$. The condition
is

\begin{equation}
\det \left\{ \left[ {\bf G}^0(\q,i\omega _m)\right] ^{-1}-
{\bf\Sigma }^{+-}(\q,i\omega _m)\right\} =0.
\end{equation}
After substituting for the matrix elements of $\left[ {\bf
G}^0(\q,i\omega _m) \right] ^{-1}$ using Eq. (25) this becomes

\begin{equation}
\left|
\begin{array}{ccc}
E _{\q,1}-i\omega _m-\frac 1\beta \Sigma _{11}^{+-}(\q,i\omega _m)
& \frac 1\beta \Sigma _{12}^{+-}(\q,i\omega _m) & \cdots \cdots \\
\frac 1\beta \Sigma _{21}^{+-}(\q,i\omega _m) & E _{\q,2}-i\omega
_m-\frac 1\beta \Sigma _{22}^{+-}(\q,i\omega _m) & \cdots \cdots
\\ \vdots & \vdots & \ddots
\end{array}
\right| =0.
\end{equation}

At low temperatures $T\ll T_C$, where the perturbative expansion
is expected to be good, an approximation to Eq. (30) (and hence to
$\Delta E$ and $\Gamma$ for each SW branch) may be obtained by
expanding the determinant to lowest order in the off-diagonal
elements. The results may eventually be expressed as

\begin{equation}
\Delta E_{\q,\nu}+i\Gamma_{\q,\nu} = -\frac 1\beta \Sigma
_{\nu\nu}^{+-}(\q, E_{\q,\nu}+i0^+) + O[\Sigma
_{\nu\nu'}^{+-}(\q,E_{\q,\nu}+i0^+)]^2,
\end{equation}
where $\nu' \not= \nu$. In deriving Eq. (31) we have assumed that
$|\Delta E+i\Gamma| \ll E$, and so the matrix elements are
evaluated there at the unrenormalized SW energy.

When the other types of self-energy terms (i.e., $\Sigma^{++}$,
$\Sigma^{--}$, and $\Sigma^{-+}$) are taken into account, it is
found that they modify Eq. (31) only by adding terms of second
order (and higher) on the right-hand side.  Therefore, to leading
order (neglecting the quadratic self-energy terms) we have

\begin{equation}
\Delta E _{\q,\nu}\simeq -\frac 1\beta \text{Re }\Sigma
_{\nu\nu}^{+-}(\q,E_{\q,\nu}+i0^+) ,
\end{equation}

\begin{equation}
\Gamma _{\q,\nu}\simeq -\frac 1\beta \text{Im }\Sigma
_{\nu\nu}^{+-}(\q,E_{\q,\nu}+i0^+),
\end{equation}
for the energy shift and damping, respectively, of SW branch $\nu$
at in-plane wave vector $\q$. In the following sections the above
expressions are used to study the effects of the three- and
four-magnon interaction processes.


\section{Three- and Four-Magnon Processes}
\subsection{Three-Magnon Processes}

The term $H^{(3)}$ in the expansion of the Hamiltonian is given in
Eq. (11). It arises entirely from the dipole-dipole interactions
and is of third order in the boson operators. It describes
splitting processes, in which one magnon is absorbed and two are
emitted, and confluence processes, in which two magnons are
absorbed and one magnon is emitted. These can lead to transfers of
energy between the different SW branches (surface and quantized
bulk modes).

The first step is to rewrite $H^{(3)}$ in terms of the new boson
operators $\alpha_{\q}^{\dagger}$ and $\alpha_{\q}$ using Eqs.
(20) and (21), so that the diagrammatic formalism may be employed.
The result is

\begin{eqnarray}
H^{(3)} &=&\frac 12\sum_{l_1l_2l_3\k\q} \left[V_1\, \alpha
_{\k,l_1}^{\dagger }\alpha _{\q-\k,l_2}^{\dagger}\alpha
_{\q,l_3}+V_2\, \alpha_{\q,l_1}^{\dagger}\alpha_{\q-\k,l_2}\alpha
_{\k,l_3}\right. \nonumber \\ &&+\left.V_3\,
\alpha_{\k,l_1}^{\dagger}\alpha_{\q-\k,l_2}^{\dagger}
\alpha_{-\q,l_3}^{\dagger}+V_4\, \alpha _{-\q,l_1}\alpha
_{\q-\k,l_2}\alpha _{\k,l_3}\right] ,
\end{eqnarray}
where the amplitude factors $V_i$ (with $i=1,2,3,4$) are listed in
Appendix B. The leading-order diagrammatic contributions to the
proper self-energy term $\Sigma_{ll}^{+-}$, as required for the
renormalization of any SW branch $l$, are shown in Fig. 5. Each of
the dotted lines represents a $V_i$ (as an ``interaction vertex'')
according to the direction of arrowing on the three Green-function
lines entering or leaving it (e.g., a $V_4$ vertex has three lines
leaving). All of the diagrams in Fig. 5 are of second order in the
vertices. Using the formal rules of diagrammatic evaluation for
interacting boson systems,$^{23}$ we obtain the self-energy
expression as

\begin{eqnarray}
\Sigma_{ll}(\q,i\omega_m)&=&-\beta
\sum_{\q'l'l''}\left\{(W_a+W_b)\frac{n^0(E_{\q',l'})} {E_{\q,
l''}}+W_c \frac{n^0(E_{\q',l'})+
n^0(E_{-\q-\q',l''})+1}{E_{-\q-\q',l''}+ E_{\q',l'}+i\omega_m }
\right. \nonumber\\ &&+ \left. W_d\frac{n^0(E_{\q',l'})+n^0(
E_{\q-\q',l''})+1}{E_{\q-\q',l''}+E_{\q',l'}
-i\omega_m}+W_e\frac{n^0(E_{\q',l'})-n^0
(E_{\q+\q',l''})}{E_{\q+\q',l''}-E_{\q',l'}- i\omega_m} \right\}.
\end{eqnarray}
The weighting factors $W_{a,b,...,e}$ are given in Appendix B and
$n^0(E) \equiv [\exp(\beta E)-1]^{-1}$ denotes the Bose-Einstein
thermal factor at energy $E$. Next, using Eqs. (32) and (33),
together with the well-known relation

\begin{equation}
\frac 1{(x\pm iE)}=\text{P}\left(\frac 1x\right) \mp i\pi
\delta(x) ,
\end{equation}
where P indicates that the principal value is taken in a summation over $x$, we obtain

\begin{eqnarray}
\Delta E_{\q ,l} &=& \text{P}
\sum_{\q'l'l''}\left\{(W_a+W_b)\frac{n^0(E_{\q',l'})} {E_{\q,
l''}}+W_c \frac{n^0(E_{\q'l'})+
n^0(E_{-\q-\q',l''})+1}{E_{-\q-\q',l''}+ E_{\q',l'}+ E_{\q ,l} }
\right. \nonumber\\ &&+ \left. W_d\frac{n^0(E_{\q'l'})+n^0(
E_{\q-\q',l''})+1}{E_{\q -\q',l''}+E_{\q',l'} -E_{\q
,l}}+W_e\frac{n^0(E_{\q',l'})-n^0 (E_{\q +\q',l''})}{E_{\q
+\q',l''}-E_{\q',l'}- E_{\q ,l}} \right\}.
\end{eqnarray}
for the energy shift, and

\begin{eqnarray}
\Gamma_{\q
,l}&=&-\pi\sum_{\q'l'l''}\left\{W_d\left[n^0(E_{\q',l'})+
n^0(E_{\q-\q',l''})+1\right]\delta(E_{\q-\q',l''}+
E_{\q',l'}-E_{\q ,l}) \right. \nonumber\\ && +
\left.W_d\left[n^0(E_{\q',l'})-n^0(E_{\q+\q',l''})
\right]\delta(E_{\q+\q',l''}-E_{\q',l'}-E_{\q ,l}) \right\},
\end{eqnarray}
for the damping. The two delta functions in Eq. (38), which
conserve energy and in-plane wave vector, correspond to the
three-magnon splitting and confluence processes respectively. The
applications of the above results for $\Delta E$ and $\Gamma$ are
discussed in Sec. VII.


\subsection{Four-Magnon Processes}

Following the previous section we first rewrite the interaction
Hamiltonian, in this case the  four-magnon contribution $H^{(4)}$
given by Eq. (13), in terms of the new boson operators. After a
straightforward but lengthy calculation we obtain the following
expression:

\begin{eqnarray}
H^{(4)} &=&\frac 12\sum_{l_1 l_2 l_3 l_4 \k\kl\q }
\left\{\Lambda_1\alpha_{\kl
l_1}^{\dagger}\alpha_{\q-\kl,l_2}^{\dagger}
\alpha_{\q-\k,l_3}\alpha_{\k,l_4}+\Lambda_2\alpha_{\q
l_1}^{\dagger}\alpha_{\q-\k-\kl,l_2}\alpha_{\kl,l_3}
\alpha_{\k,l_4} \right.\nonumber\\ &&+
\Lambda_3\alpha_{\kl_1}^{\dagger}\alpha_{\kl,l_2}^{\dagger}
\alpha_{\q-\k-\kl,l_3}^{\dagger}\alpha_{\q,l_4}+\Lambda_4
\alpha_{\kl l_1}^{\dagger}\alpha_{\q-\kl,l_2}^{\dagger}
\alpha_{\k-\q,l_3}^{\dagger}\alpha _{-\k,l_4}^{\dagger
}\nonumber\\ &&\left.+\Lambda_5\alpha_{-\kl
l_1}\alpha_{\kl-\q,l_2} \alpha_{\q-\k,l_3}\alpha_{\k,l_4}\right\},
\end{eqnarray}
where the amplitude factors  $\Lambda_i$ ($i=1,2,3,4$) associated
with each operator term are given in Appendix C. These terms
define the interaction vertices of the diagrammatic formalism. By
contrast to the three-magnon case, the four-magnon interactions
involve dipolar and exchange terms and contribute even in the
Heisenberg limit (see, e.g., Kontos and Cottam$^{15,16}$ for the
case of a semi-infinite Heisenberg ferromagnet).

According to Eqs. (32) and (33), the formal results for the SW
energy shift and damping related to $H^{(4)}$ arise from the real
and imaginary parts, respectively, of an appropriately-chosen self
energy term in the $\Lambda_i$ vertices. The leading-order
contribution (denoted by $\Sigma^{(1)}$) to the proper self-energy
$\Sigma^{+-}(\q,i\omega _m)$ is of first order in $\Lambda_i$ and
corresponds to the diagram in Fig. 6a, together with topologically
similar diagrams. Again the dotted line represents an interaction
vertex corresponding to the number of lines entering or leaving.
On evaluating the diagrams we obtain the contribution to the
energy shift of SW branch $l$ as

\begin{eqnarray}
\Delta E_{\q ,l}&=&\sum_{\q'l'}\Theta_a{n^0(E_{\q',l'})}
\end{eqnarray}
where the weighting term $\Theta_a$ arises from a combination of
$\Lambda_1$ vertices and is defined in Appendix C.  It is the
analog, for the thin film with exchange and dipolar coupling, of
the Dyson SW interaction term for the case of bulk SW in an
infinite Heisenberg ferromagnet (see, e.g., Ref. 12).

It turns out that the self-energy part $\Sigma^{(1)}$ is real, and
so it gives a vanishing contribution to the damping.  Therefore,
to calculate the damping, it is necessary to consider the
self-energy contributions $\Sigma^{(2)}$ that are of {\it second}
order in the $\Lambda_i$ vertices.  This corresponds to diagrams
with a more complicated loop structure (and hence with more
internal variables). In Fig. 6b we show just those diagrams for
$\Sigma^{(2)}$ that give rise to a damping term (i.e., those that
have an imaginary part after the analytic continuation $i\omega_m
\rightarrow E_{\q ,l}+i0^+$ is made. The final result for the
damping of SW branch $l$ is

\begin{eqnarray}
\Gamma_{\q ,l}&=&-\pi\sum_{\q'\q''
l'l''l'''}\left\{\Theta_b\left\{n^0(E_{\q'-\q ,l'})
\left[{n^0(E_{\q'',l''})}+1\right]-{n^0(E_{\q'-\q'',l'''})}
\nonumber\right.\right.\\
&&\times\left.\right[n^0(E_{\q'',l''})-n^0(E_{\q'-\q,
l'})\left]\right\} \delta\left(E_{\q'',l''}+E_{\q'-\q'',l'''}-
E_{\q'-\q ,l'}-E_{\q, l}\right)\nonumber\\
&&+\Theta_c\left\{n^0(E_{\q'',l''})n^0(E_{\q'-\q'',l'''})+
\left[n^0(E_{\q'',l''})+1\right]\right.\nonumber\\
&&\times\left.\left[n^0(E_{\q'',l''})+n^0(E_{\q'-\q'',l'''})+1\right]\right\}
\delta\left(E_{\Q'',l''}+E_{\q-\q',l'}+E_{\q'-\q'',l'''}- E_{\q
,l}\right)\nonumber\\ &&+\Theta_d\left\{n^0(E_{\q'-\q
,l'})\times\left[n^0(E_{\q'',l''}) +1\right]-n^0(E_{\q'-\q
,l'})\right.\nonumber\\
&&\times\left.\left.\right[n^0(E_{\q'',l''})-n^0(E_{\q'-\q
,l'})\left]\right\}
\delta\left(E_{\q'+\q'',l'''}-E_{\q'',l''}-E_{\q'-\q ,l'}-E_{\q
,l}\right)\right\}.
\end{eqnarray}
The second-order weighting factors $\Theta_{b,c,d}$, which
multiply the three different delta functions conserving energy and
in-plane wave vector, are given in Appendix C.


\section{Numerical Results and Discussion}

Numerical calculations for particular materials are now presented
based on the formal expressions for $\Delta E_{\q ,l}$ and
$\Gamma_{\q ,l}$ in the previous section. This provides us
with predictions of the overall dependence on $\q$ and $T$ for the
different SW branches and allows us to compare the relative
importance of the three- and four-magnon processes in various
situations.

We start by considering ultrathin films of EuO, where the dipolar
interactions have a significant effect on the long-wavelength
linear SW properties and the exchange effects dominate at shorter
wavelengths (see Sec. III). We assume the same values for $4\pi M$
and $H_{Ex}$ as before, but for simplicity we ignore the
next-nearest exchange $J_2$. In Fig. 7a we show the damping of the
lowest SW branch (labeled 1) for $N = 8$ and for temperature $T\ll
T_C$ as a function of the in-plane wave-vector component $q_x$.
The mode under consideration is the analog of the DE surface mode
at small $q_x$ which is then modified by the exchange at larger
$q_x$. In this case it is found that the dominant damping
mechanism comes from the three-magnon term $H^{(3)}$ in second
order, i.e., from Eq. (38). The delta functions in Eq. (38) give
$E_{\q ,l} = E_{\q \mp \q',l''} \pm E_{\q',l'}$, in accordance
with conservation of energy and in-plane wave vector. For a SW
with in-plane wave vector $\q$ and branch $l$, these correspond
physically to the splitting (upper signs) and confluence (lower
signs) processes. Here the labels $l'$ and $l''$ may be the same
as, or different from, $l$. Along with the total three-magnon
damping contribution (solid curve) we show in Fig. 7a the
contributions from some of individual $(l',l'')$ processes. It is
seen that several of the inter-branch processes (i.e., where at
least one of $l'$ and $l''$ differs from $l$) play an important
role. In Fig. 7b we make some comparisons between the total
damping for the SW branch 1 with $N = 8$ (solid curve) and $N =16$
(dotted line). In this particular case the damping has a similar
behavior for both values of $N$, although the individual
inter-branch contributions (not shown) are different. In the same
plot we show the total damping for branch 2 in the case of $N=8$
(dashed curve).  This is slightly larger than the damping for
branch 1, mainly because there are SW branches above and below it,
giving more possibilities to satisfy the required conservation
conditions.

It is interesting to compare the damping results predicted for EuO
with those for other materials. We consider Fe where the ratio of
dipolar to exchange strength is smaller than in EuO, as well as
GdCl$_3$ where the ratio is larger. For the case of a GdCl$_3$
film with $N=8$ layers, the total damping at $T\ll T_C$ is shown
in Fig. 8, where we use the same $4\pi M$ and $H_{Ex}$ as in Sec.
III. This presents a rather different behavior, particularly for
small $q_x$, when compared to a EuO film (see Fig. 7a) and also to
a Fe film (see Fig. 9). The parameters used for the latter case
are $4\pi M = 2.14$ T and $H_{Ex}$ = 2140 T. The qualitative
differences for the dominant three-magnon damping in the case of
GdCl$_3$ is mainly because the influence of the dipolar terms
(including the role of the DE surface mode) extends over a greater
range of wave vectors than is the case for EuO or Fe. This makes
it easier to satisfy the energy and wave-vector conservation
conditions for the damping at smaller wave vectors.

As mentioned earlier, the three-magnon damping has contributions
due to splitting and confluence processes. The temperature
dependence of these processes arise from the combinations of Bose
factors associated with each delta-function term. We note, in
particular, that the second term (the confluence term) on the
right-hand side of Eq. (38) vanishes as $T \rightarrow 0$. This is
as expected since there needs to be a thermally-excited SW to
participate in the confluence. By contrast, the first term in Eq.
(38) does not vanish in the zero-temperature limit, since it is
always possible (provided the conservation conditions are
satisfied) for a SW to split into two modes. The numerical
calculations presented above were all for the low-temperature
limit where the three-magnon splitting is dominant. In Fig. 10 we
show calculations for the temperature dependence of the
three-magnon damping, taking the case of a EuO film (with the same
parameters as in Fig. 7a) and a fixed value of the wave-vector
component ($q_x/\pi=0.5$). It can be seen that, at low enough
temperatures, the splitting process (solid curve) dominates and
increases slowly with temperature, whereas the confluence process
(dashed curve) increases very rapidly with temperature.

The four-magnon damping, which is given by Eq. (41), contains
three different processes proportional to the weighting factors
$\Theta_b$, $\Theta_c$ and $\Theta_d$, together with their Bose
factors. The delta functions provide for conservation of energy
and in-plane wave vectors. The only non-vanishing contribution as
$T \rightarrow 0$ is the term proportional to $\Theta_c$, which
describes a splitting of an incoming SW into {\it three} modes. It
is proportional to the dipole-dipole sums (i.e., it vanishes in
the exchange limit) and is generally less important than the
splitting process in Eq. (38) discussed above. The other two terms
in Eq. (41) contribute to the four-magnon damping when $T \not=
0$. The first term (proportional to $\Theta_b$) is the analog of
the scattering term in the damping for Heisenberg
systems.$^{12,15,16}$ The incoming SW scatters off a
thermally-excited SW into two other SW modes. The term
proportional to $\Theta_d$ corresponds to the incoming SW
scattering off a pair of SW modes into another SW mode. It arises
specifically as a consequence of the dipolar terms in the
Hamiltonian.

We now turn to a discussion of results for the SW energy shifts,
which are given by Eqs. (37) and (40) for the three- and
four-magnon cases respectively. In a low-temperature limit $T\ll
T_C$ the dominant contributions come from the three-magnon
processes, specifically from the terms proportional to $W_c$ and
$W_d$ in Eq. (37). Numerical results for $\Delta E_{\q ,l}$ versus
$q_x$ for EuO films in the same three cases as Fig. 7b are shown
in Fig. 11. The main difference (compared to the damping) is that
the inter-branch terms are much more important. This arises
basically as a consequence of $\Delta E_{\q ,l}$ and $\Gamma_{\q
,l}$ coming from the real and imaginary parts, respectively, of
self-energy terms in the diagram formalism, and these may have
quite different dependencies. For example, the imaginary parts
have delta functions, which lead to peaks and structural features
in the damping, whereas the summations in the real parts tend to
give a smoother behavior for the SW energy shifts. Hence $\Delta
E$ has a weaker dependence on $q_x$ in Fig. 11 than was the case
for the damping. The larger $\Delta E$ value and its
$q_x$-dependence for the thicker film (where the linear SW modes
are closer together) is due to the inter-branch terms in the
summations.

At low but finite temperatures with $T\ll T_C$ the four-magnon
contribution to $\Delta E_{\q ,l}$ begins to play a role,
particularly for the lower-energy SW branches, because the Bose
factor in Eq. (40) becomes nonzero. Some numerical calculations
are shown in Figs. 12 and 13 for EuO films with $N = 8$. In Fig.
12 the four-magnon contribution to the energy shift for the lowest
SW branch ($l=1$) is plotted as a function of wave-vector
component $q_x$ at a fixed temperature ($T=0.04$ K). It is seen
that the qualitative behavior is quite different from that for the
three-magnon contribution (see Fig.11). We note also that the
$q_x$-dependence in Fig. 12 is different from that of the
four-magnon energy shift for a Heisenberg
ferromagnet,$^{12,15,16}$ which is proportional to $q_x^2$ at
small wave vectors. This difference, compared to Heisenberg
systems, is partly due to the the dipolar terms in the factor
$\Theta_a$ in Eq. (40) and partly due to the dipolar contribution
to the SW energy gap. In Fig. 13 we compare the temperature
dependence of the three-magnon SW energy shift (solid curve) with
that due to four-magnon processes (dashed curve). The latter
quantity increases rapidly with temperature and eventually
dominates. This is again for a EuO film, but for a fixed wave
vector corresponding to $q_x =0.5\pi$.

\section{CONCLUSIONS}

We have developed a microscopic (Hamiltonian-based) theory for the
dipole-exchange SW modes and the interactions between them in
ultrathin ferromagnetic films. We first used the theory to study
the linear dispersion relation of the discrete SW in EuO and
GdCl$_3$, including the effects of next-nearest neighbor exchange.
The results obtained with the microscopic theory are in good
agreement with those of the usual macroscopic theories in limiting
cases where there are a large number of layers and the wave
vectors are very small. Outside of these regimes it becomes
necessary to use a microscopic approach, as in this work.

Secondly, the theory enabled us to develop the leading-order
three- and four-magnon interaction terms in the Hamiltonian for
the film. These three- and four-magnon processes can involve modes
from the different discrete SW branches (surface or bulk) in all
combinations. Formal expressions were found for the energy shift
and damping of each branch.  Applications were made to EuO, Fe,
and  GdCl$_3$. In the case where the three-magnon processes
dominate, it has been shown that there are significantly different
types of behavior predicted for GdCl$_3$ compared to EuO or Fe.
This happens because the influence of the dipolar terms extends
over a greater range of wave vectors in the GdCl$_3$. As a
consequence, the energy and wave-vector conservation conditions
for the damping are more readily satisfied at smaller wave
vectors. Our method enabled us to carry out calculations also at
higher temperatures, where the four-magnon processes play a more
significant role. The above-mentioned influence of the dipolar
interactions also makes an important contribution to these
processes. The major effect, compared to the four-magnon energy
shift in a Heisenberg ferromagnet without dipolar interactions,
can be observed in the $q_x$ dependence of the four-magnon energy
shift.

In order to probe the effects discussed here, inelastic light
scattering would be an appropriate experimental technique.$^{19}$
An interesting extension of our work would be to apply the
nonlinear theory  in an examination of SW instabilities in
ultrathin films under conditions of parallel or perpendicular
``pumping'' by a microwave field.

\acknowledgments

The authors gratefully acknowledge partial support from NSERC
(Canada) and the Brazilian agencies CNPq, CAPES, FINEP, and
FUNCAP.


\appendix
\section{DIPOLE SUMS}

Here we list the expressions and properties of the dipole sums
required for this work. They are calculated as in Ref. 8.

The sums of dipole interactions between spins located in {\it
different} atomic layers (i.e., with $y \equiv \| (n-n')\| \neq0$)
have the following expressions:

\begin{equation}
D_{n,n^{\prime }}^{xx}\left( q_x,q_z\right) =4\pi
\sum_{l,m=-\infty }^\infty \frac{\left( \pi l+q_x/2\right)
^2}{\gamma_{lm}}\exp \left( -2\left| y\right|\gamma_{lm}\right),
\end{equation}

\begin{equation}
D_{n,n^{\prime }}^{yy}\left( q_x,q_z\right) =-4\pi
\sum_{l,m=-\infty }^\infty \gamma_{lm}\ \exp \left( -2\left|
y\right| \gamma_{lm}\right),
\end{equation}

\begin{equation}
D_{n,n^{\prime }}^{zz}\left( q_x,q_z\right) =4\pi
\sum_{l,m=-\infty }^\infty \frac{\left( \pi m+q_z/2\right)
^2}{\gamma_{lm}}\exp \left( -2\left| y\right| \gamma_{lm}\right),
\end{equation}

\begin{equation}
D_{n,n^{\prime }}^{xy}\left( q_x,q_z\right) =-i4\pi
sgn(y)\sum_{l,m=-\infty }^\infty \left( \pi l+q_x/2\right) \exp
\left( -2\left| y\right| \gamma_{lm}\right),
\end{equation}

\begin{equation}
D_{n,n^{\prime }}^{zy}\left( q_x,q_z\right) =-i4\pi
sgn(y)\sum_{l,m=-\infty }^\infty \left( \pi m+q_z/2\right) \exp
\left( -2\left| y\right| \gamma_{lm}\right),
\end{equation}

\begin{equation}
D_{n,n^{\prime }}^{xz}\left( q_x,q_z\right) =4\pi
\sum_{l,m=-\infty }^\infty \frac{\left( \pi l+q_x/2\right) \left(
\pi m+q_z/2\right) }{\gamma_{lm}}\exp \left( -2\left| y\right|
\gamma_{lm}\right).
\end{equation}
In the expressions above we denote $\gamma_{lm}=[(\pi l+q_x/2)^2+ (\pi
m+q_z/2)^2]$, and $K_i(x)$ is a modified Bessel function of integer order
$i$ for any variable $x$.

For the dipolar sums involving spins in the {\it same}
layer ($y=0$) we have the following expressions for the diagonal
terms (involving the same superscript):

\begin{equation}
D_{n,n}^{xx}\left( q_x,q_z\right) =-2S_x\left( q_x,q_y\right)
+S_z\left( q_x,q_y\right),
\end{equation}

\begin{equation}
D_{n,n}^{yy}\left( q_x,q_z\right) =S_x\left( q_x,q_y\right)
+S_z\left( q_x,q_y\right),
\end{equation}

\begin{equation}
D_{n,n^{}}^{zz}\left( q_x,q_z\right) =-2S_z\left( q_x,q_y\right)
+S_x\left( q_x,q_y\right),
\end{equation}
where

\begin{equation}
S_x\left( q_x,q_y\right) =\frac{16}3\sum_{x=1}^\infty
\sum_{m=-\infty }^\infty \left( \pi m+q_z/2\right) ^2\cos \left(
q_xx\right) \left[ K_2\left( 2x\left| \pi m+q_z/2\right| \right)
\right],
\end{equation}

\begin{equation}
S_z\left( q_x,q_y\right) =\frac{16}3\sum_{z=1}^\infty
\sum_{m=-\infty }^\infty \left( \pi m+q_x/2\right) ^2\cos \left(
q_zz\right) \left[ K_2\left( 2z\left| \pi m+q_x/2\right| \right)
\right].
\end{equation}
Also for the off-diagonal terms:

\begin{equation}
D_{n,n}^{xz}\left( q_x,q_z\right) =16\sum_{x=1}^\infty
\sum_{m=-\infty }^\infty \left( \pi m+q_z/2\right) \left| \pi
m+q_z/2\right| \sin\left( q_xx\right)\left[ K_1\left( 2x\left| \pi
m+q_z/2\right| \right)\right],
\end{equation}
while $D_{n,n}^{xy}\left(q_x,q_z\right)=D_{n,n}^{yz}\left(
q_x,q_z\right)=0 $ by symmetry. Finally, when $q_x = q_z =0$, we
have
\begin{equation} D_{n,n}^{zz}\left( 0,0\right) =-\frac{4\pi
^2}9\left[ 1+24\sum_{m=1}^\infty m^2K_2\left( 2x\pi m\right)
\right].
\end{equation}

\section{Three-magnon Amplitude Factors}

The expressions for the amplitude factor $V_i$ ($i=1,2,3,4$) in
Eq. (34), and hence for the vertices in the diagrammatic
representation, are given by

\begin{eqnarray}
V_1(\k ,\q|l_1,l_2,l_3)
&=&\sum_{nn'}\left\{A_{nn'}^{(3)}(\k)S_{n'l_1}^*(\k)
S_{nl_2}^*(\q-\k)S_{nl_3}(\q)\right. \nonumber\\ &&+
A_{nn'}^{(3)}(\k)S_{n'l_1}^*(\k)T_{nl_2}^*(\k-\q)T_{nl_3}(\q)\nonumber\\
&&+
A_{nn'}^{(3)}(-\k)T_{n'l_1}^*(-\k)S_{nl_2}^*(\q-\k)S_{nl_3}(\q)\nonumber\\
&&+
A_{nn'}^{(3)}(-\k)T_{n'l_1}^*(-\k)T_{nl_2}^*(\k-\q)T_{nl_3}(-\q)\nonumber\\
&&+
A_{nn'}^{(3)}(-\q)T_{nl_1}^*(-\k)S_{nl_2}^*(\q-\k)T_{n'l_3}(-\q)\nonumber\\
&&\left.+
A_{nn'}^{(3)}(\q)S_{nl_1}^*(\k)T_{nl_2}^*(\k-\q)S_{n'l_3}(\q)\right\},
\end{eqnarray}

\begin{eqnarray}
V_2(\k ,\q|l_1,l_2,l_3)
&=&\sum_{nn'}\left\{A_{nn'}^{(3)}(\k)S_{nl_1}^*(\q)
S_{nl_2}(\q-\k)S_{n'l_3}(\k)\right. \nonumber\\ &&+
A_{nn'}^{(3)}(\k)T_{nl_1}^*(-\q)T_{nl_2}(\k-\q)S_{n'l_3}(\k)\nonumber\\
&&+
A_{nn'}^{(3)}(-\k)S_{nl_1}^*(\q)S_{nl_2}(\q-\k)T_{n'l_3}(-\k)\nonumber\\
&&+
A_{nn'}^{(3)}(-\k)T_{nl_1}^*(-\q)T_{nl_2}(\k-\q)T_{n'l_3}(-\k)\nonumber\\
&&+
A_{nn'}^{(3)}(-\q)T_{n'l_1}^*(-\q)S_{nl_2}(\q-\k)T_{nl_3}(-\k)\nonumber\\
&&\left.+
A_{nn'}^{(3)}(\q)S_{n'l_1}^*(\q)T_{nl_2}(\k-\q)S_{nl_3}(\k)\right\},
\end{eqnarray}

\begin{eqnarray}
V_3(\k
,\q|l_1,l_2,l_3)&=&\sum_{nn'}\left\{A_{nn'}^{(3)}(\k)S_{n'l_1}^*(\k)
S_{nl_2}^*(\q-\k)T_{nl_3}^*(\q)\right.\nonumber\\ &&\left.+
A_{nn'}^{(3)}(-\k)T_{n'l_1}^*(-\k)T_{nl_2}^*(\k-\q)S_{nl_3}^*(-\q)\right\},
\end{eqnarray}

\begin{eqnarray}
V_4(\k
,\q|l_1,l_2,l_3)&=&\sum_{nn'}\left\{A_{nn'}^{(3)}(\k)T_{nl_1}(\q)
S_{nl_2}(\q-\k)S_{n'l_3}(\k)\right.\nonumber\\ &&\left.+
A_{nn'}^{(3)}(-\k)S_{nl_1}(-\q)T_{nl_2}(\k-\q)
T_{n'l_3}(-\k)\right\}.
\end{eqnarray}

In terms of the above quantities in Eqs. (35), (37) and (38), we
can now define the weighting factors $W_i$ ($i=a,b,c,d,e$)  as

\begin{eqnarray}
W_a &=&\left[V_1\left(0, \q'|l'',l',l'\right)+V_1\left(\q',\q'|
l',l'',l'\right)\right] \left[V_2\left(0,\q|l,l,l''
\right)+V_2\left(\q,\q|l,l'',l \right) \right] ,
\end{eqnarray}

\begin{eqnarray}
W_b &=&\left[V_2\left(0,\q'|l'',l',l'\right)+
V_2\left(\q',\q'|l',l'',l' \right)\right] \left[V_1\left( 0,\q|
l,l,l'' \right)+V_1\left(\q,\q|l,l'',l\right) \right],
\end{eqnarray}

\begin{eqnarray}
W_d
&=&\left[V_1\left(\q',\q|l',l'',l\right)+V_1\left(\q-\q',\q|l'',l',l
\right)\right]\nonumber\\ && \times
\left[V_2\left(\q',\q|l',l'',l\right)+V_2\left(\q-\q',\q|l'',l',l
\right)\right],
\end{eqnarray}

\begin{eqnarray}
W_e &=&\left[V_1\left(\q',\q+\q'|l',l,l''\right)+
V_1\left(\q,\q+\q'|l,l',l''\right)\right]\nonumber\\&& \times
\left[ V_2\left(\q',\q+\q'|l'',l,l'
\right)+V_2\left(\q,\q+\q'|l'',l',l \right)\right],
\end{eqnarray}
and finally $W_c \equiv w_1 w_2+w_2W_{c2}+w_1W_{c3}+2W_{c4}$, where

\begin{eqnarray}
w_1 = V_3\left(\q',-\q|l',l'',l\right)+
V_3\left(-\q-\q',-\q|l'',l',l \right),
\end{eqnarray}

\begin{eqnarray}
w_2 = V_4\left(\q',-\q|l,l'',l'\right)+
V_4\left(-\q-\q',-\q|l,l',l''\right),
\end{eqnarray}

\begin{eqnarray}
W_{c2} &=& V_3\left(\q,-\q'|l,l'',l'\right)+
V_3\left(-\q-\q',-\q'|l'',l,l' \right) \nonumber\\&&
 + V_3\left(\q,\q+\q'|l,l',l''\right)+ V_3\left(\q',\q+\q'|l',l,l''
\right),
\end{eqnarray}

\begin{eqnarray}
W_{c3} &=& V_4\left(\q,-\q'|l',l'',l\right)+
V_4\left(-\q-\q',-\q'|l',l,l''\right) \nonumber\\ &&
+V_4\left(\q,\q+\q'|l'',l',l\right)+
V_4\left(\q',\q+\q'|l'',l,l'\right),
\end{eqnarray}

\begin{eqnarray}
W_{c4} &=&\left[V_3\left(\q',-\q|l',l'',l\right)+
V_3\left(-\q-\q',-\q|l'',l',l\right)+\right.\nonumber\\
&&\left.V_3\left(\q',\q+\q|l',l,l''\right)+
V_3\left(\q,\q+\q'|l,l',l''\right)\right]\nonumber\\ && \times
\left[V_4\left(\q',-\q|l,l'',l'\right)+
V_4\left(-\q-\q',-\q'|l,l',l'' \right)+\right.\nonumber\\ &&\left.
V_4\left(\q',\q+\q'|l'',l,l'\right)+ V_4\left(\q,\q+\q'|l'',l',l
\right)\right].
\end{eqnarray}


\section{Four-magnon Amplitude Factors}

The expressions for the amplitude factor $\Lambda_i$ ($i=1,2,3$)
in Eq. (39), and hence for the vertices in the diagrams in Fig. 6,
are given by

\begin{eqnarray}
\Lambda_1(\k\k'\q|l_1l_2l_3l_4)& &=\sum_{nn'}\left\{A_{nn'}^{(4)}(
\k)\left[S_{n,l_1}^{*}
(\kl)S_{n,l_2}^{*}(\q-\kl)S_{n,l_3}(\q-\k)S_{n',l_4}(\k)+
2S_{n,l_1}^{*}(\kl)\times \right.\right. \nonumber\\
&&T_{n,l_2}^{*}(\kl-\q)T_{n,l_3}(\k-\q)S_{n',l_4}(\k)+
T_{n,l_1}^{*}(-\kl)T_{n,l_2}^{*}(\k-\q)
T_{n,l_3}(\kl-\q)\times\nonumber\\
&&\left.T_{n',l_4}(-\k)+2T_{n,l_1}^{*}(-\kl)S_{n,l_2}^{*}(\q-\kl)
S_{n,l_3}(\q-\k)T_{n',l_4}(-\k)\right]+
A_{n,n'}^{(4)}(\kl)\times\nonumber\\
&&\left[T_{n',l_1}^{*}(-\kl)T_{n,l_2}^{*}(\k-\q)T_{n,l_3}(\kl-\q)
T_{n,l_4}(-\k)+2T_{n',l_1}^{*}(-\kl)\times \right.\nonumber\\
&&S_{n,l_2}^{*}(\q-\kl)S_{n,l_3}(\q-\k)T_{n,l_4}(-\k)+
S_{n',l_1}^{*}(\kl)S_{n,l_2}^{*}(\q-\kl)\times\nonumber\\
&&\left.S_{n,l_3}(\q-\k)S_{n,l_4}(\k)+2S_{n'',l_1}^{*}(\kl)
T_{n,l_2}^{*}(\kl-\q)T_{n,l_3}(\k-\q)S_{n,l_4}(\k)\right]-\nonumber\\
&&2A_{n,n'}^{\prime(4)}(\k-\kl)\left[S_{n',l_1}^{*}(\kl)
S_{n,l_2}^{*}(\q-\kl)S_{n,l_3}(\q-\k)S_{n',l_4}(\k)+\right.\nonumber\\
&&T_{n',l_1}^{*}(-\kl)T_{n,l_2}^{*}(\k-\q)T_{n,l_3}(\kl-\q)
T_{n',l_4}(-\k)+S_{n',l_1}^{*}(\kl)T_{n,l_2}^{*}(\kl-\q)\times\nonumber\\
&&\left.T_{n,l_3}(\k-\q)S_{n',l_4}(\k)+T_{n',l_1}^{*}(-\kl)
S_{n,l_2}^{*}(\q-\kl)S_{n,l_3}(\q-\k)T_{n',l_4}(-\k)\right]-\nonumber\\
&&2A_{n,n'}^{\prime(4)}(\q)\left[S_{n',l_1}^{*}(\kl)
T_{n',l_2}^{*}(\kl-\q)S_{n,l_3}(\q-\k)T_{n,l_4}(-\k)+
T_{n,l_1}^{*}(-\kl)\times\right.\nonumber\\
&&\left.S_{n,l_2}^{*}(\q-\kl)
T_{n',l_3}(\k-\q)S_{n',l_4}(\k)\right]+B_{n,n'}^{(4)}(\k)
\left[T_{n,l_1}^{*}(-\kl)T_{n,l_2}^{*}(\kl-\q)\times\right.\nonumber\\
&&\left. T_{n,l_3}(\k-\q)S_{n',l_4}(\k)+2S_{n,l_1}^{*}(\kl)
T_{n,l_2}^{*}(\kl-\q)S_{n,l_3}(\q-\k)S_{n',l_4}(\k)\right]+\nonumber\\
&&B_{n,n'}^{(4)}(-\kl)\left[T_{n',l_1}^{*}(-\kl)S_{n,l_2}^{*}(\q-\kl)
S_{n,l_3}(\q-\k)S_{n,l_4}(\k)+2T_{n',l_1}^{*}(-\kl)\times\right.\nonumber\\
&&\left.T_{n,l_2}^{*}(\kl-\q)S_{n,l_3}(\q-\k)T_{n,l_4}(-\k)\right]
+C_{n,n'}^{(4)}(\kl)\left[S_{n',l_1}^{*}(\kl)
T_{n,l_2}^{*}(\kl-\q)\right.\times\nonumber\\
&&\left.T_{n,l_3}(\k-\q)T_{n,l_4}(-\k)+2S_{n',l_1}^{*}(\kl)
S_{n,l_2}^{*}(\q-\kl)T_{n,l_3}(\k-\q)S_{n,l_4}(\k)\right]+\nonumber\\
&&C_{n,n'}^{(4)}(-\k)\left[S_{n,l_1}^{*}(\kl)S_{n,l_2}^{*}(\q-\kl)
S_{n,l_3}(\q-\k)T_{n',l_4}(-\k)+ \right.\nonumber \\
&&\left.\left.2S_{n,l_1}^{*}(\kl)T_{n,l_2}^{*}(\kl-\q)
T_{n,l_3}(\k-\q)T_{n',l_4}(-\k)\right]\right\} ,
\end{eqnarray}

\begin{eqnarray}
\Lambda _2(\k\k'\q|l_1l_2l_3l_4)& &
=\sum_{n,n'}\left\{A_{n,n'}^{(4)}(\k)
\left[T_{n,l_1}^{*}(-\q)T_{n,l_2}(\k+\kl-\q)T_{n,l_3}(-\kl)
S_{n',l_4}(\k)+2S_{n,l_1}^{*}(\q)\times \right. \right.\nonumber\\
&&T_{n,l_2}(\k+\kl-\q)S_{n,l_3}(\kl)S_{n',l_4}(\k)+
S_{n,l_1}^{*}(\q)S_{n,l_2}(\q-\k-\kl)S_{n,l_3}(\kl)\times\nonumber\\
&&\left.T_{n',l_4}(-\k)+2T_{n,l_4}^{*}(-\q)T_{n,l_2}(\k+\kl-\q)
S_{n,l_3}(\kl)T_{n',l_4}(-\k)\right]+A_{n,n'}^{(4)}(\q)\times\nonumber\\
&&\left[T_{n',l_1}^{*}(-\q)T_{n,l_2}(\k+\kl-\q)S_{n,l_3}(\kl)
T_{n,l_4}(-\k)+S_{n',l_1}^{*}(\q)T_{n,l_2}(\k+\kl-\q)\times\right.\nonumber\\
&&\left.S_{n,l_3}(\kl)S_{n',l_4}(\k)\right]-2A_{n,n'}^{\prime(4)}(\k-\q)
\left[S_{n',l_1}^{*}(\q)T_{n,l_2}(\k+\kl-\q)S_{n,l_3}(\kl)+\right.\nonumber\\
&&\left.S_{n',l_4}(\k)+T_{n',l_1}^{*}(-\q)T_{n,l_2}(\k+\kl-\q)
S_{n,l_3}(\kl)T_{n',l_4}(-\k)\right]-\nonumber\\
&&2A_{n,n'}^{\prime(4)}(\k+\kl)\left[S_{n,l_1}^{*}(\q)
S_{n,l_2}(\q-\k-\kl)T_{n',l_3}(-\kl)S_{n',l_4}(\k)+\right.\nonumber\\
&&\left.T_{n,l_1}^{*}(-\q)T_{n,l_2}(\k+\kl-\q)T_{n',l_3}(-\kl)
S_{n',l_4}(\k)\right]+B_{n,n'}^{(4)}(\k)
\left[S_{n,l_1}^{*}(\q)\right.\times\nonumber\\
&&S_{n,l_2}(\q-\k-\kl)S_{n,l_3}(\kl)S_{n',l_4}(\k)+
2T_{n,l_1}^{*}(-\q)S_{n,l_2}(\q-\k-\kl)\nonumber\\
&&\left.T_{n,l_3}(-\kl)S_{n',l_4}(\k)\right]+
C_{n,n'}^{(4)}(-\k)\left[T_{n,l_1}^{*}(-\q)
T_{n,l_2}(\k+\kl-\q)\times\right.\nonumber\\
&&T_{n,l_3}(-\kl)T_{n',l_4}(-\k)+2S_{n,l_1}^{*}(\q)T_{n,l_2}(\k+\kl-\q)
S_{n,l_3}(\kl)\times\nonumber\\
&&\left.T_{n',l_4}(-\k)\right]+B_{n,n'}^{(4)}(-\q)\left[
T_{n',l_1}^{*}(-\q)S_{n,l_2}(\q-\k-\kl)S_{n,l_3}(\kl)\times\right.\nonumber\\
&&\left.\left.T_{n,l_4}(-\k)\right]+C_{n,n'}^{(4)}(\q)\left[S_{n',l_1}^{*}(\q)
T_{n,l_2}\left( \k+\kl-q\right)
T_{n,l_3}(-\kl)S_{n,l_4}(\k)\right]\right\},
\end{eqnarray}

\begin{eqnarray}
\Lambda _3(\k\k'\q|l_1l_2l_3l_4)& &=\sum_{n,n'}\left\{
A_{n,n'}^{(4)}(-\k)
\left[T_{n',l_1}^{*}(-\k)S_{n,l_2}^{*}(\k')S_{n,l_3}^{*}(\q-\k-\k')
S_{n,l_4}(\q)+\right.\right.\nonumber\\
&&2T_{n',l_1}^{*}(-\k)S_{n,l_2}^{*}(\kl)T_{n,l_3}^{*}(\k+\kl-\q)
T_{n,l_4}(-\q)+S_{n',l_1}^{*}(\k)\times\nonumber\\
&&T_{n,l_2}^{*}(-\kl)T_{n,l_3}^{*}(\k+\kl-\q)T_{n,l_4}(-\q)
+2S_{n',l_1}^{*}(\k)S_{n,l_2}^{*}(\kl)\times\nonumber\\
&&\left.T_{n,l_3}^{*}(\k+\kl-\q)S_{n,l_4}(\q)\right]
+A_{n,n'}^{(4)}(\q)\left[S_{n,l_1}^{*}(\k)S_{n,l_2}^{*}(\kl)\times\right.\nonumber\\
&&T_{n,l_3}^{*}(\k+\kl-\q)S_{n',l_4}(\q)+T_{n,l_1}^{*}(-\k)
T_{n,l_2}^{*}(-\kl)S_{n,l_3}^{*}(\q-\k-\kl)\times\nonumber\\
&&\left.T_{n',l_4}(-\q)\right]-2A_{n,n'}^{\prime(4)}(\k-\q)
\left[S_{n',l_1}^{*}(\k)S_{n,l_2}^{*}(\kl)
T_{n,l_3}^{*}(\k+\kl-\q)\times\right.\nonumber\\
&&\left.S_{n',l_4}(\q)+T_{n',l_1}^{*}(-\k)T_{n,l_2}^{*}(-\kl)
S_{n,l_3}^{*}(\q-\k-\kl)T_{n',l_4}(-\q)\right]-\nonumber\\
&&2A_{n,n'}^{\prime(4)}(\k+\kl)\left[T_{n',l_1}^{*}(-\k)
S_{n',l_2}^{*}(\kl)S_{n,l_3}^{*}(\q-\k-\kl)
S_{n,l_4}(\q)+\right.\nonumber\\
&&\left.T_{n',l_1}^{*}(-\k)S_{n',l_2}^{*}(\kl)T_{n,l_3}^{*}(\k+\kl-\q)
T_{n,l_4}(-\q)\right]+B_{n,n'}^{(4)}(-\k)\times\nonumber\\
&&\left[T_{n',l_1}^{*}(-\k)T_{n,l_2}^{*}(-\kl)T_{n,l_3}^{*}(\k+\kl-\q)
T_{n,l_4}(-\q)+2T_{n',l_1}^{*}(-\k)\times\right.\nonumber\\
&&\left.S_{n,l_2}^{*}(\kl)T_{n,l_3}^{*}(\k+\kl-\q)S_{n,l_4}(\q)\right]+
C_{n,n'}^{(4)}(\k)\left[S_{n',l_1}^{*}(\k)S_{n,l_2}^{*}(\kl)\times\right.\nonumber\\
&&S_{n,l_3}^{*}(\q-\k-\kl)S_{n,l_4}(\q)+2S_{n',l_1}^{*}(\k)
S_{n,l_2}^{*}(\kl)T_{n,l_3}^{*}(\k+\kl-\q)\times\nonumber\\
&&\left.T_{n,l_4}(-\q)\right]+B_{n,n''}^{(4)}(\q)
\left[S_{n,l_1}^{*}(\k)T_{n,l_2}^{*}(-\kl)T_{n,l_3}^{*}(\k+\kl-\q)S_{n',l_4}(\q)\right].\nonumber\\
&&\left.+C_{n,n'}^{(4)}(-\q)\left[T_{n,l_1}^{*}(-\k)
S_{n,l_2}^{*}(\kl)S_{n,l_3}^{*}(\q-\k-\kl)T_{n',l_4}(-\q)\right]\right\}.
\end{eqnarray}
The terms $\Lambda_4$ and $\Lambda_5$ are given by similar
expressions. They will not be quoted here, since they do not enter
into the expressions for the SW energy shift and damping.

We can now define the weighting factors $\Theta_i$ ($i=a,b,c,d$) in Eqs. (40) and (41) as

\begin{eqnarray}
\Theta_a&=&\Lambda_1(\q,\q,\q+\q'|l,l',l',l)+\Lambda_1(\q,\q',\q+\q'|l',l,l',l)\nonumber\\
&+&
\Lambda_1(\q',\q,\q+\q'|l,l',l,l')+\Lambda_1(\q',\q',\q+\q',|l',l,l,l'),
\end{eqnarray}

\begin{eqnarray}
\Theta_b &=&\left[\Lambda_1(\q,\q'',\q'|l''l'''l'l)+
\Lambda_1(\q,\q'-\q'',\q'|l'''l''l'l)+\right. \nonumber\\
&&\left.\Lambda_1(\q'-\q,\q'',\q'|l''l'''ll') +
\Lambda_1(\q'-\q,\q'-\q'',\q'|l'''l''ll')\right]\times\nonumber\\
& & \left[\Lambda_1(\q'',\q,\q'|ll'l'''l'')+
\Lambda_1(\q'',\q'-\q,\q'|l'll'''l'')+\right.\nonumber \\
&&\left.\Lambda_1(\q'-\q'',\q,\q'|ll'l''l''')+
\Lambda_1(\q'-\q'',\q'-\q,\q'|l'll''l''')\right],
\end{eqnarray}

\begin{eqnarray}
\Theta_c &=&\left[\Lambda_2(\q-\q',\q'',\q|ll'''l''l')+
\Lambda_2(\q-\q',\q'-\q'',\q|ll''l'''l')+\right.\nonumber\\
&&\Lambda_2(\q'',\q-\q',\q|ll'''l'l'')+
\Lambda_2(\q'',\q'-\q'',\q|ll'l'''l'')+\nonumber\\
&&\left.\Lambda_2(\q'-\q'',\q'',q|ll'l''l''')+
\Lambda_2(\q'-\q'',\q-\q',\q|ll''l'l''')\right]\times \nonumber\\
&&\left[\Lambda_3(\q-\q',\q'',\q|l'l''l'''l)+
\Lambda_3(\q-\q',\q'-\q'',\q|l'l'''l''l)+\right.\nonumber\\
&&\Lambda_3(\q'',\q-\q',\q|l''l'l'''l)+
\Lambda_3(\q'',\q'-\q'',\q|l''l'''l'l)+\nonumber\\
&&\left.\Lambda_3(\q'-\q'',\q'',\q|l'''l''l'l)+
\Lambda_3(\q'-\q','\q-\q',\q|l'''l'l''l)\right] ,
\end{eqnarray}

\begin{eqnarray}
\Theta_d &=&\left[\Lambda_2(\q,\q'',\q'+\q''|l'''l'l''l)+
\Lambda_2(\q,\q'-\q,\q'+\q''|l'''l''l'l)\right.+\nonumber\\
&&\Lambda_2(\q'',\q,\q'+\q''|l'''l'll'')+
\Lambda_2(\q'',\q'-\q,\q'+\q''|l'''ll'l'')+\nonumber\\
&&\left.\Lambda_2(\q'-\q,\q,\q'+\q''|l'''l''ll')+
\Lambda_2(\q'-\q,\q'',\q'+\q''|l'''ll''l')\right]\times
\nonumber\\ &&\left[\Lambda_3(\q,\q'',\q'+\q''|ll''l'l''')+
\Lambda_3(\q,\q'-\q,\q'+\q''|ll'l''l''')\right.+\nonumber\\
&&\Lambda_3(\q'',\q,\q'+\q''|l''ll'l''')+
\Lambda_3(\q'',\q-\q,\q'+\q''|l''l'll''')+\nonumber\\
&&\left.\Lambda_3(\q'-\q,\q,\q'+\q''|l'll''l''')+
\Lambda_3(\q'-\q,\q'',\q'+\q''|l'l''ll''')\right] .
\end{eqnarray}



\begin{figure}

\begin{figure}
\caption{The effects of the next-nearest neighbor exchange
interactions on the SW spectrum in a 16-layer EuO film: (a) SW
frequency against $J_2/J_1$ for $q_x=0.001$ (solid curve) and
$q_x=0.04$ (dashed curve); (b) SW frequency against $q_x$ for
$J_2/J_1=0$ (solid curve), $J_2/J_1=0.2$ (dashed curve), and
$J_2/J_1=0.3$ (dot-dashed curve).}

\end{figure}

\begin{figure}
\caption{The linear SW dispertion relation for a 16-layer GdCl$_3$
film with next-nearest neighbor exchange $J_2=0.25J_1$ solid lines
and $J_2=0$ dashed lines.}
\end{figure}

\begin{figure}
\caption{Diagrammatic representation of (a) the non-interacting
Green function and (b) the possible types of proper self-energy
contributions of the system.}
\end{figure}

\begin{figure}
\caption{Diagrammatic representation of the Dyson series for the
renormalized Green function in terms of the self-energy parts.}
\end{figure}

\begin{figure}
\caption{Second order diagrams for the self-energy
contributions due to three-magnon processes.}
\end{figure}

\begin{figure}
\caption{Self-energy diagrams due to four-magnon
processes in (a) first order and (b) second order.}
\end{figure}

\begin{figure}
\caption{Three-magnon damping of the lowest SW branch (1)
vs reduced wave vector $q_xa/\pi$ for EuO films: (a) Total
damping (solid curve) and contributions from some of individual
$(l',l'')$ processes: (1,1) dot-dashed curve,(1,2)+(2,1) dashed
curve, and (2,2) dotted curve; (b) Total damping for the cases
of $N = 8$ (solid curve),$N =16$ (dotted line) for the lowest
branch 1, and for $N=8$ and branch 2 (dashed curve).}
\end{figure}

\begin{figure}
\caption{ Three-magnon damping for a GdCl$_3$ film
with 8 layers for SW branch 1.}
\end{figure}

\begin{figure}
\caption{As in Fig. 8, but for a Fe film. }
\end{figure}

\begin{figure}
\caption{Temperature dependence of the three-magnon damping for
a 8-layer EuO film with a fixed value of the wave vector
($q_x/\pi=0.5$), showing (a) splitting processes (solid
curve), b) confluence processes (dashed curve).}
\end{figure}

\begin{figure}
\caption{SW energy shift vs reduced wave vector $q_xa/\pi$ for EuO
films with $N=8$ (solid curve) and $N=16$ (dotted curve) for the
lowest branch 1, and $N=8$ for branch 2 (dashed curve).}
\end{figure}

\begin{figure}
\caption{The four-magnon SW energy shift of the lowest branch vs
reduced wave vector $q_xa/\pi$ for an EuO film with 8 layers.}
\end{figure}

\begin{figure}
\caption{Temperature dependence of the SW energy shift compared
for three-magnon (solid curve) and four-magnon (dashed curve)
processes.}
\end{figure}

\end{figure}

\newpage
\centerline{\epsfysize=13cm\epsfbox{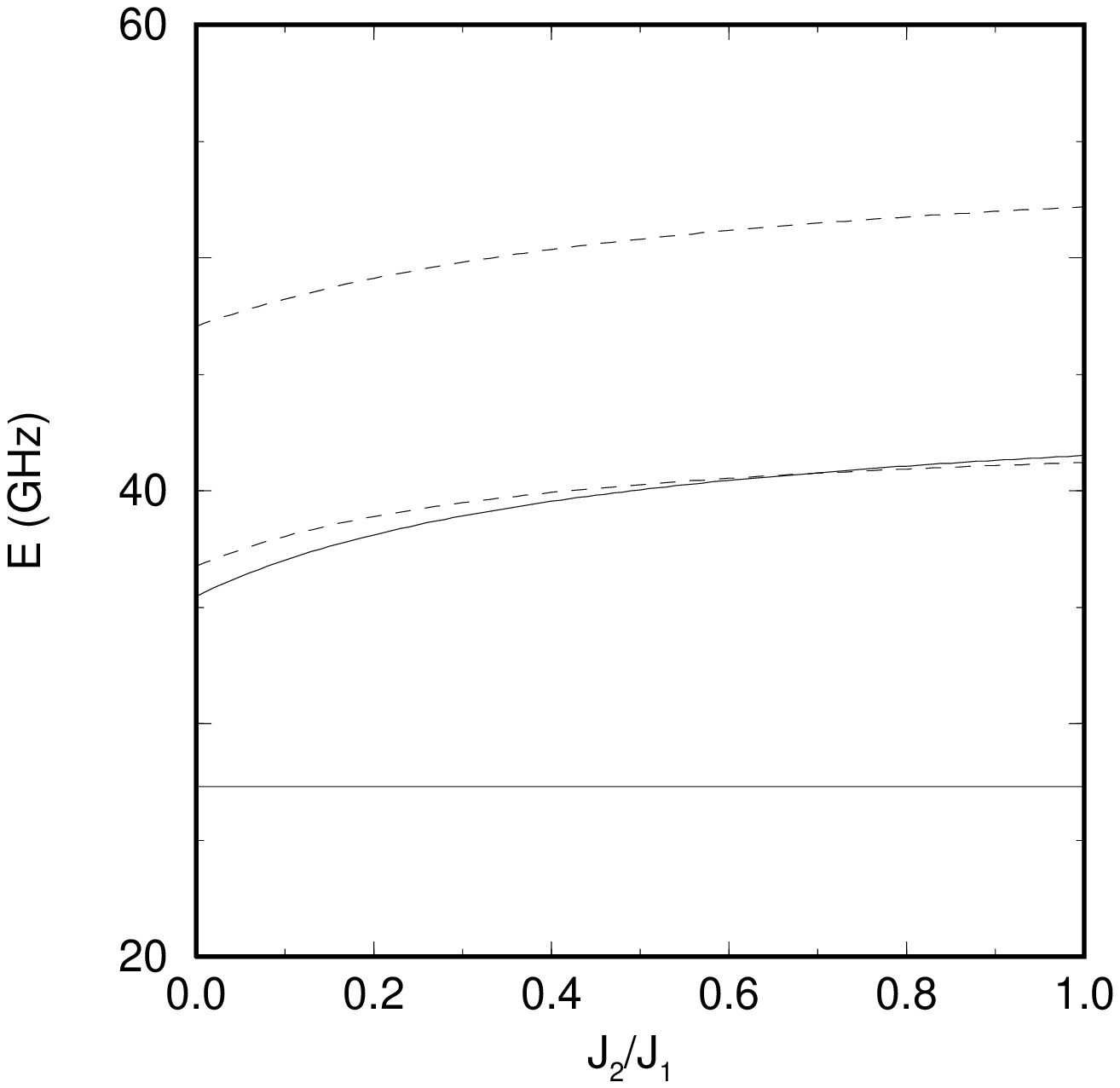}} \vskip5cm
\centerline {Fig. 1a.}

\centerline{\epsfysize=13cm\epsfbox{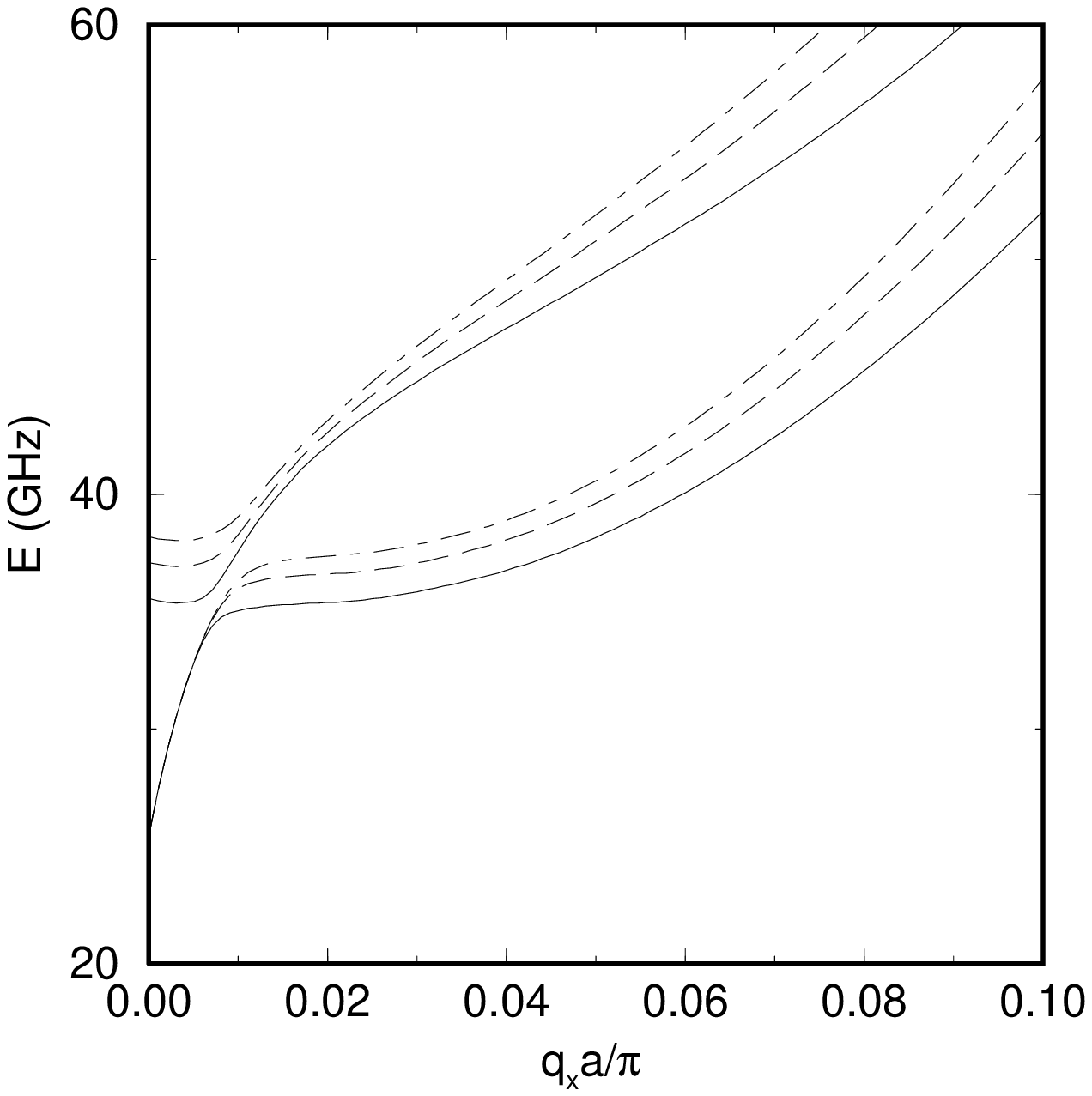}} \vskip5cm
\centerline {Fig. 1b.}

\centerline{\epsfysize=13cm\epsfbox{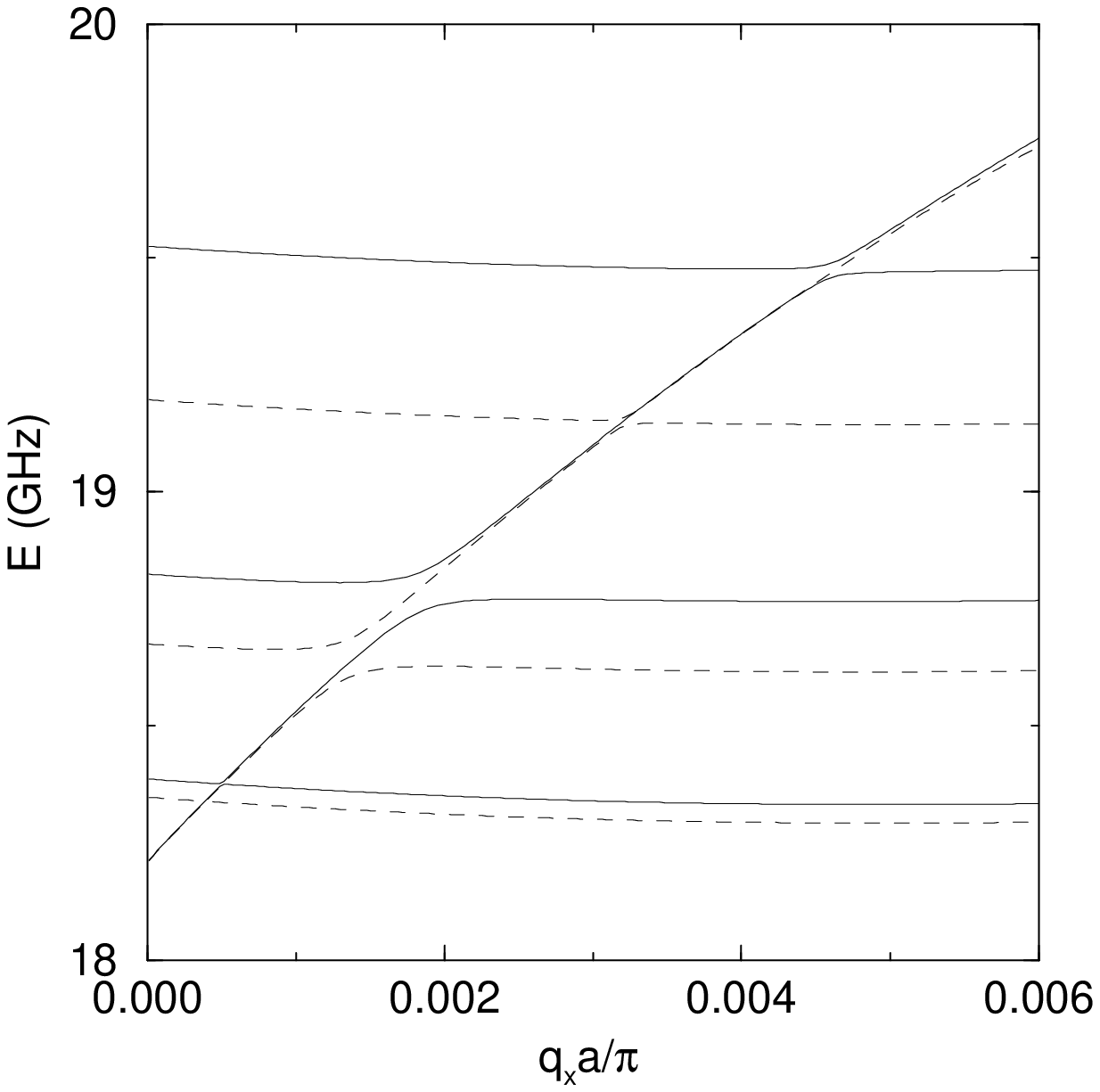}} \vskip5cm
\centerline {Fig. 2.}

\newpage

\centerline{\epsfysize=5cm \epsfbox{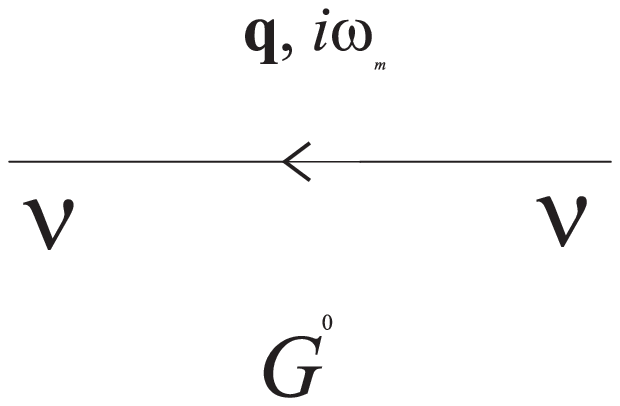}}\vskip2cm
\centerline {Fig.3a.}

\vskip 3cm \centerline{\epsfysize=8cm\epsfbox{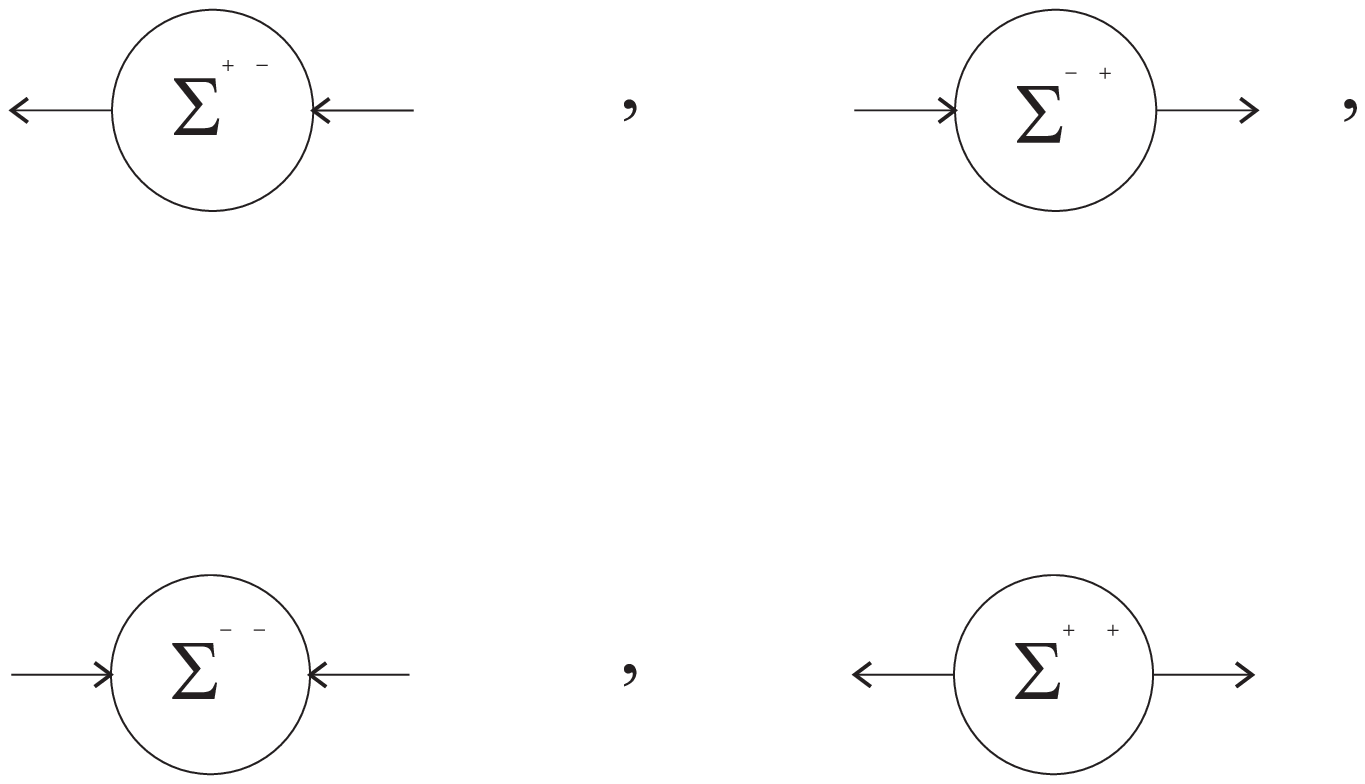}}
\vskip2cm \centerline {Fig. 3b.}
\newpage
\centerline{\epsfysize=13cm\epsfbox{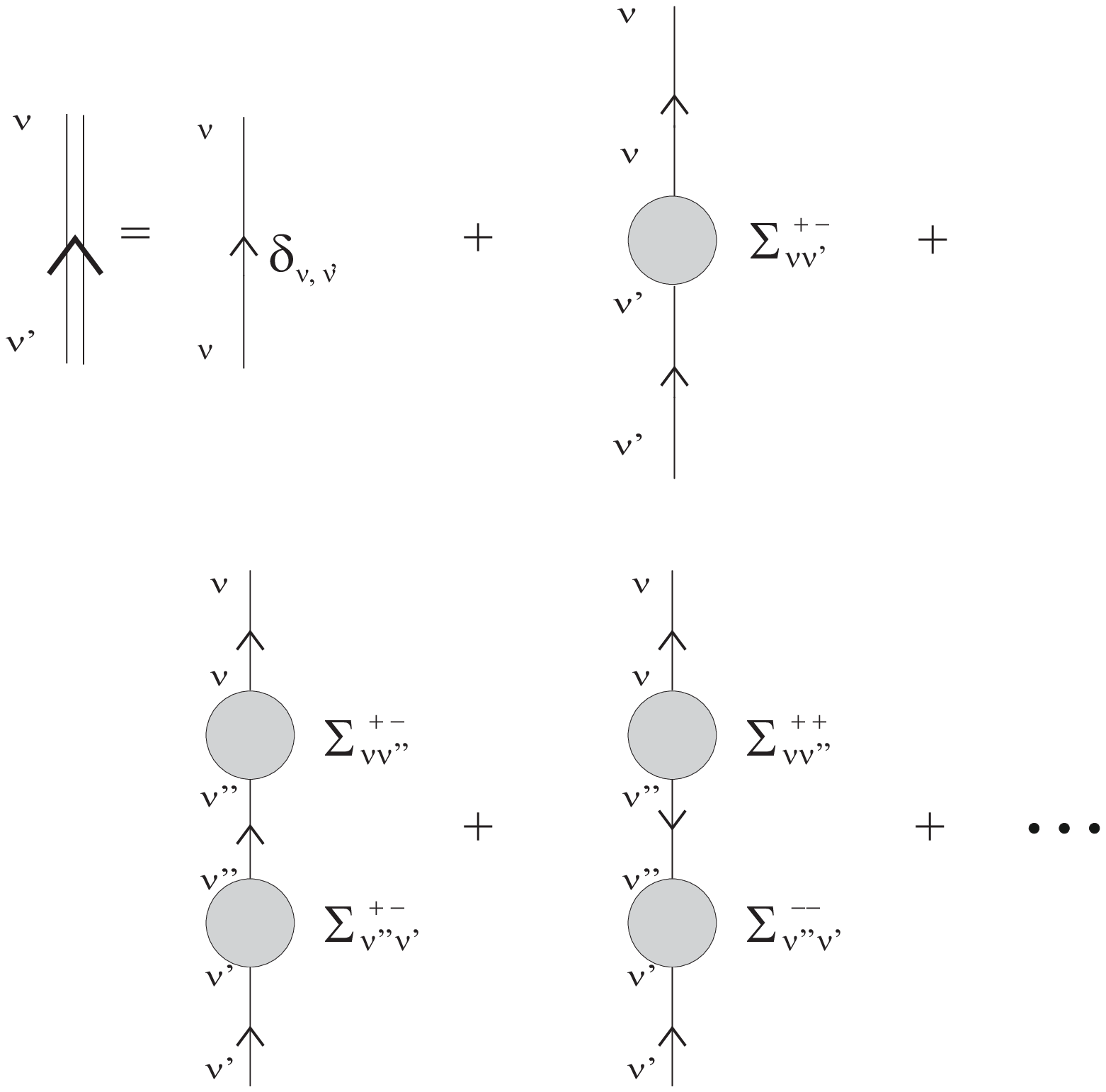}} \vskip5cm
\centerline {Fig. 4.}

\centerline{\epsfysize=13cm\epsfbox{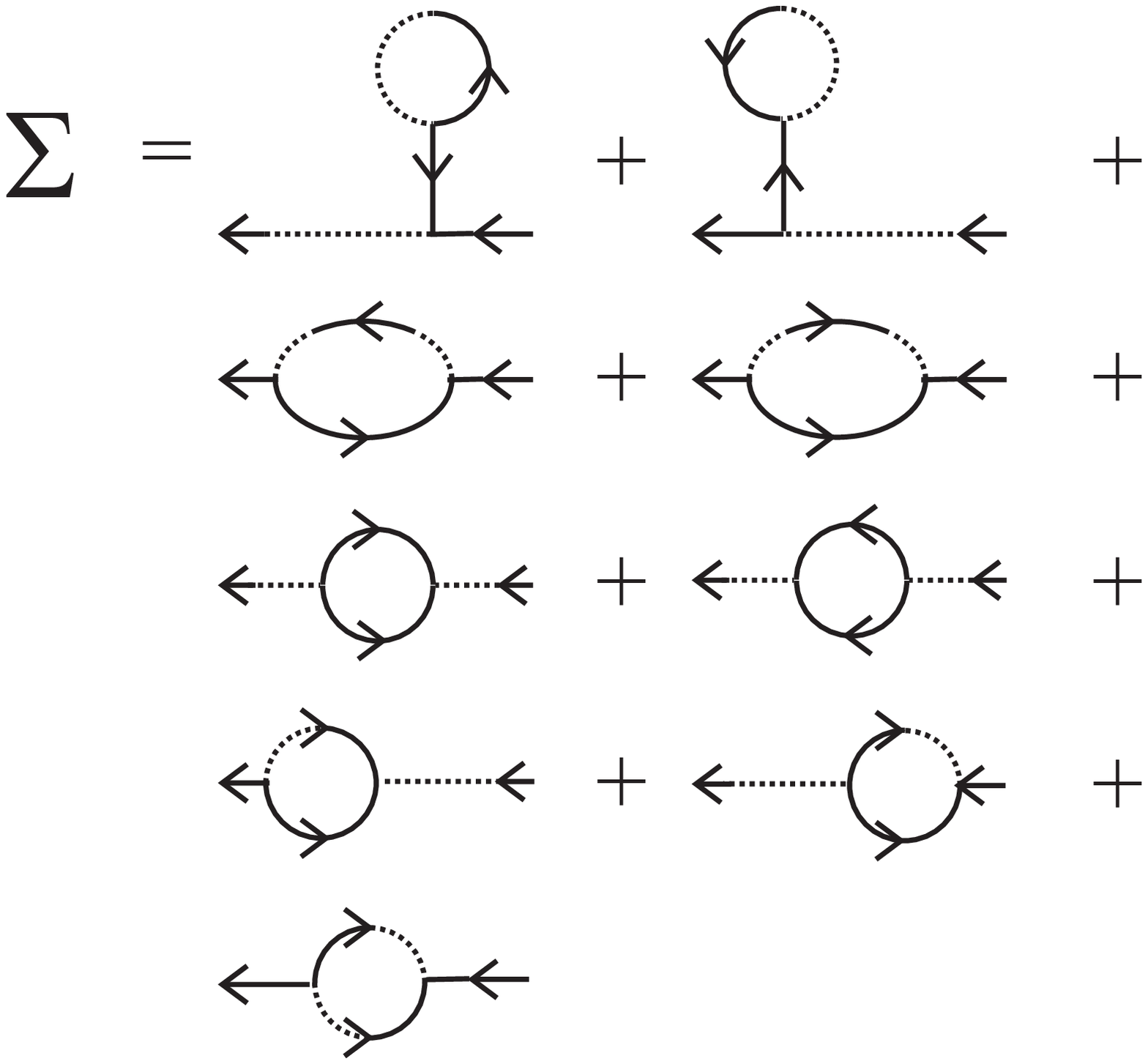}} \vskip5cm
\centerline {Fig. 5.}
\newpage
\centerline{\epsfysize=3cm\epsfbox{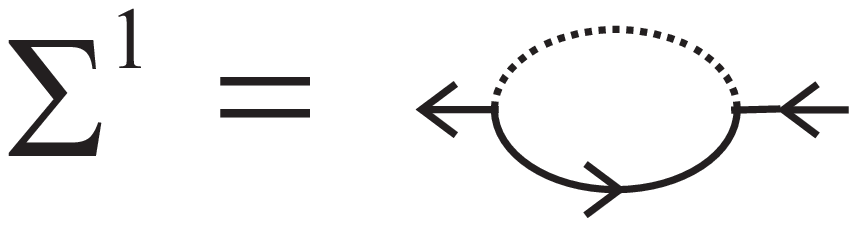}} \vskip5cm
\centerline {Fig. 6a.} \vskip 2cm
\centerline{\epsfysize=6cm\epsfbox{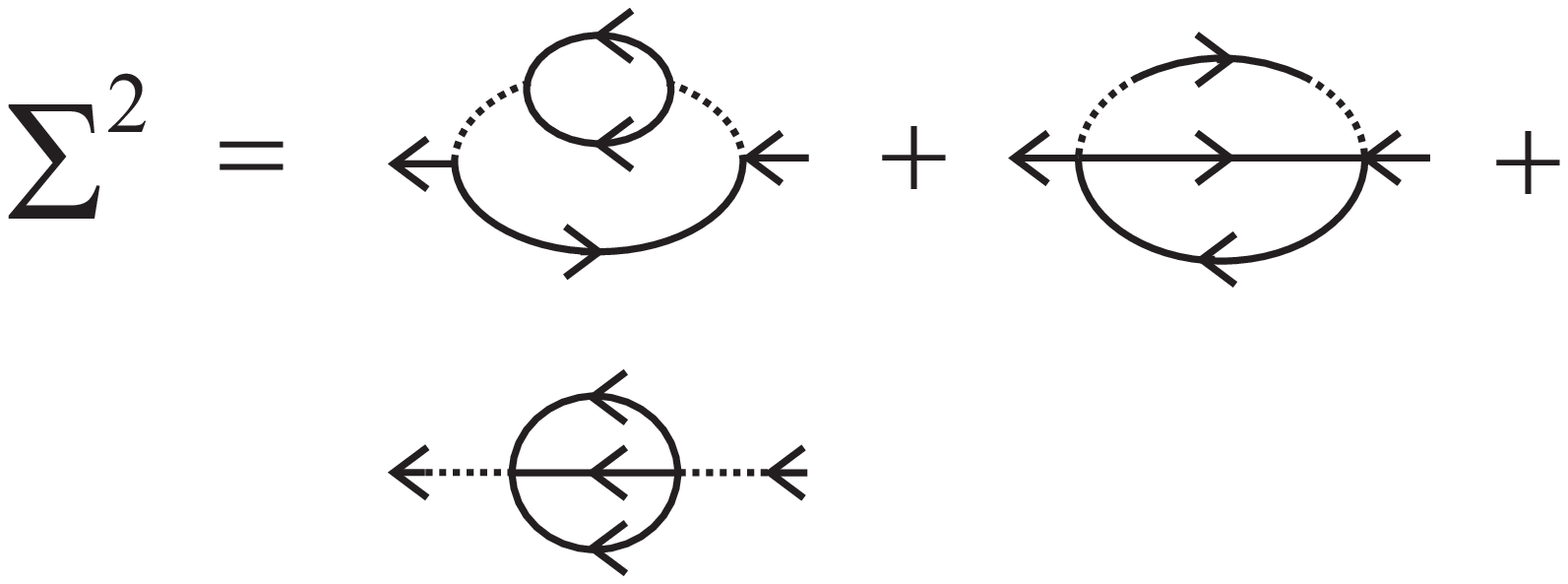}} \vskip5cm
\centerline {Fig. 6b.}
\centerline{\epsfysize=12cm\epsfbox{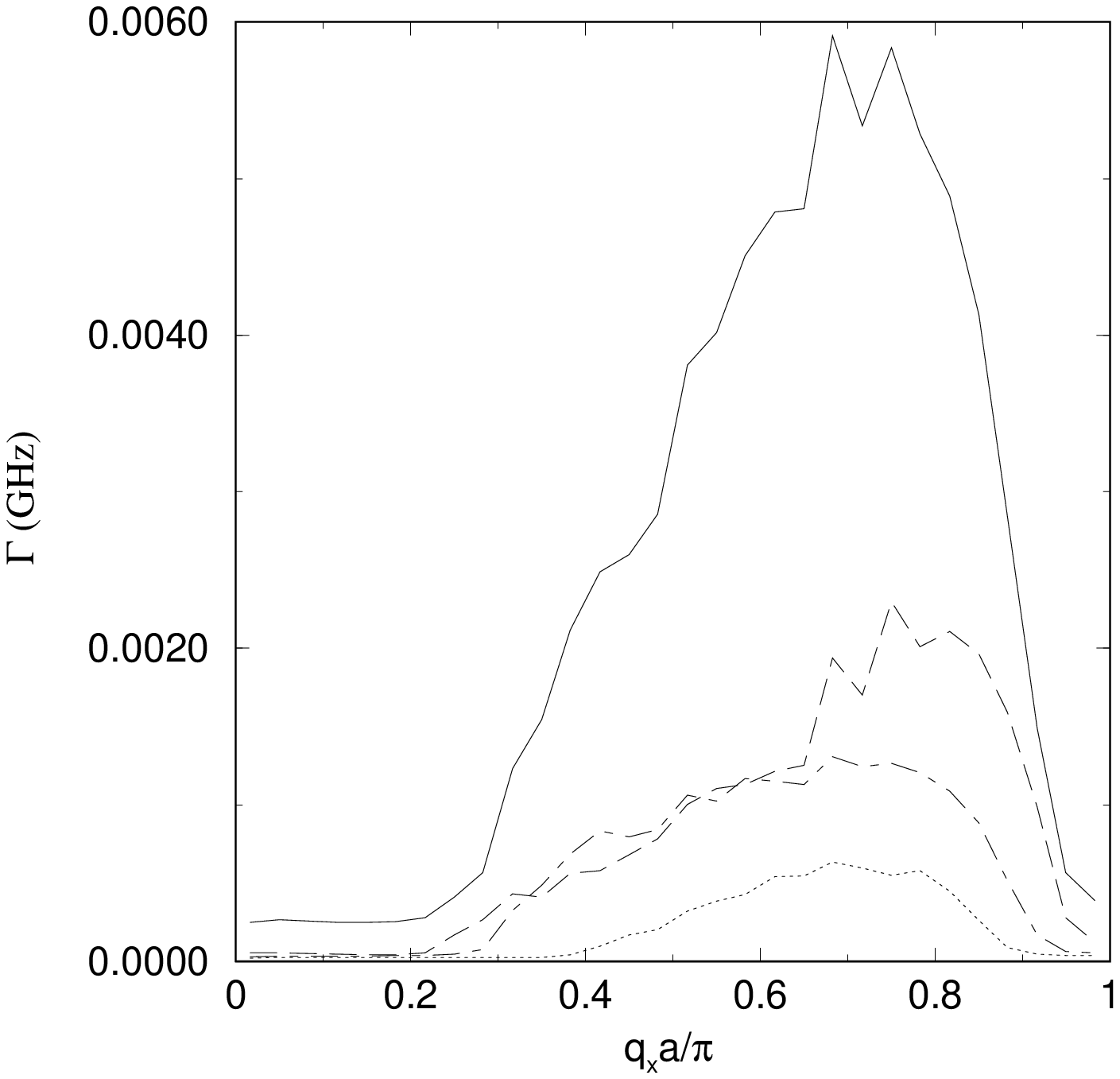}} \vskip5cm
\centerline {Fig. 7a.}
\centerline{\epsfysize=12cm\epsfbox{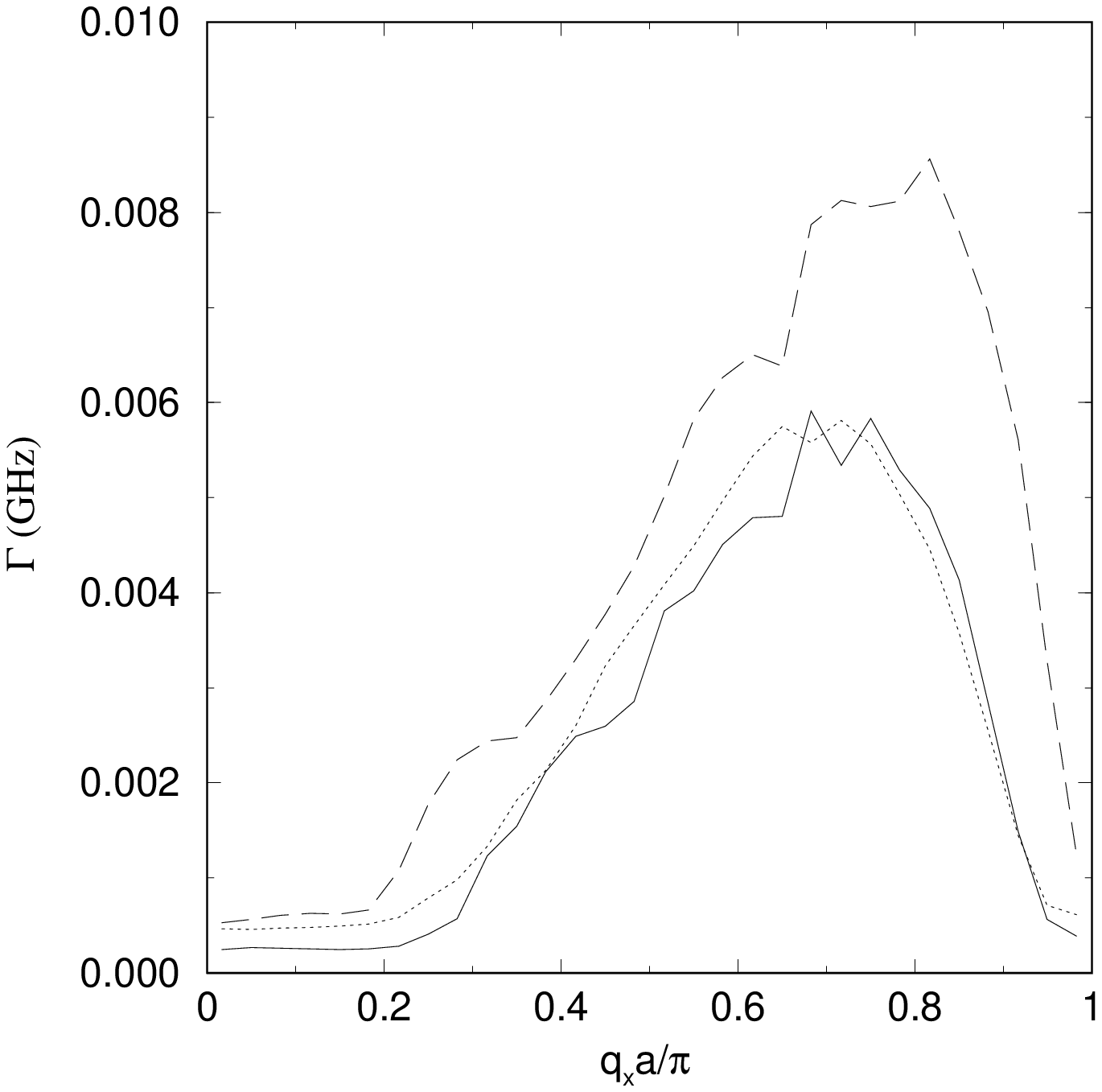}} \vskip5cm
\centerline {Fig. 7b.}
\centerline{\epsfysize=12cm\epsfbox{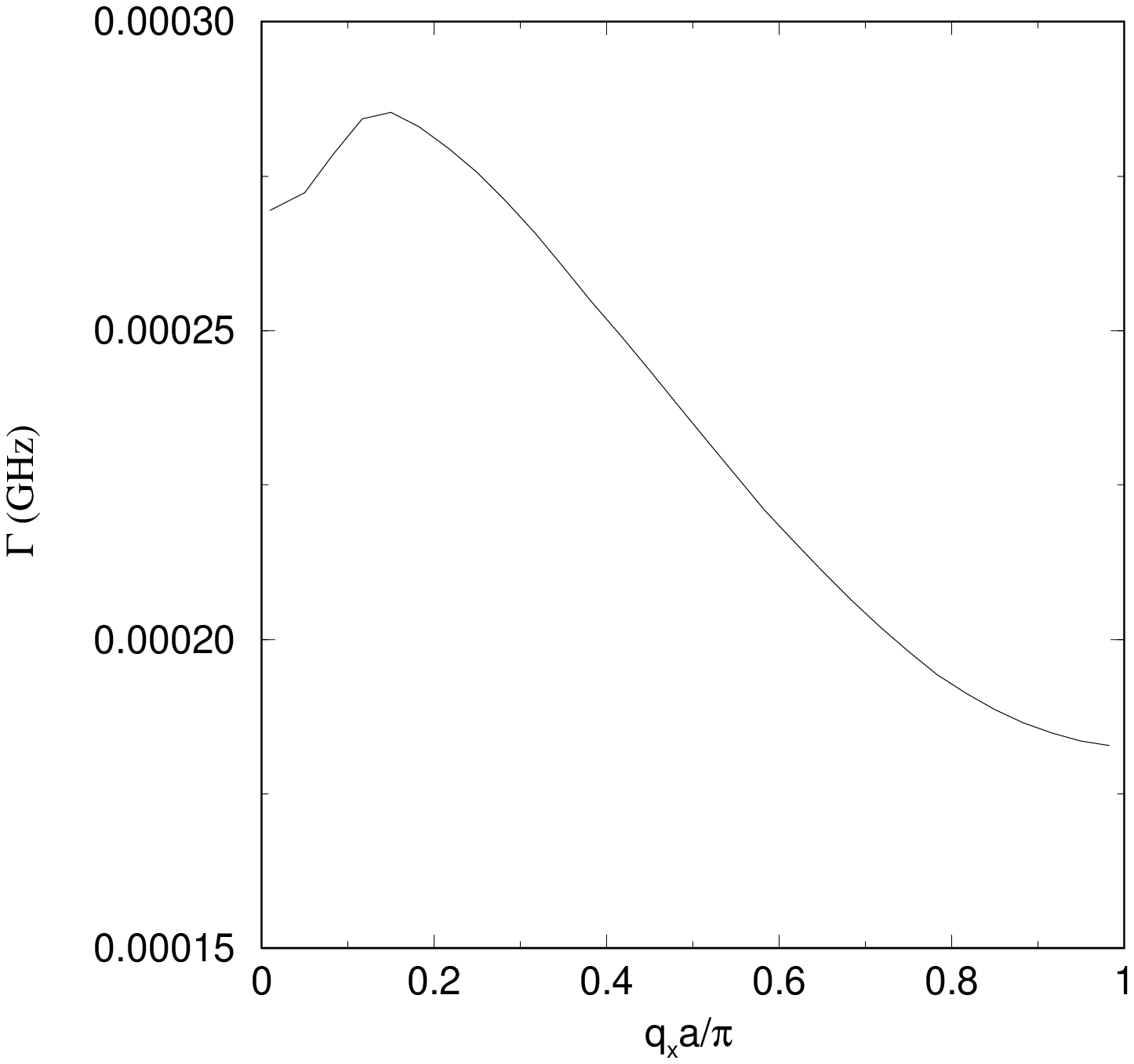}} \vskip5cm
\centerline {Fig. 8.}
\centerline{\epsfysize=12cm\epsfbox{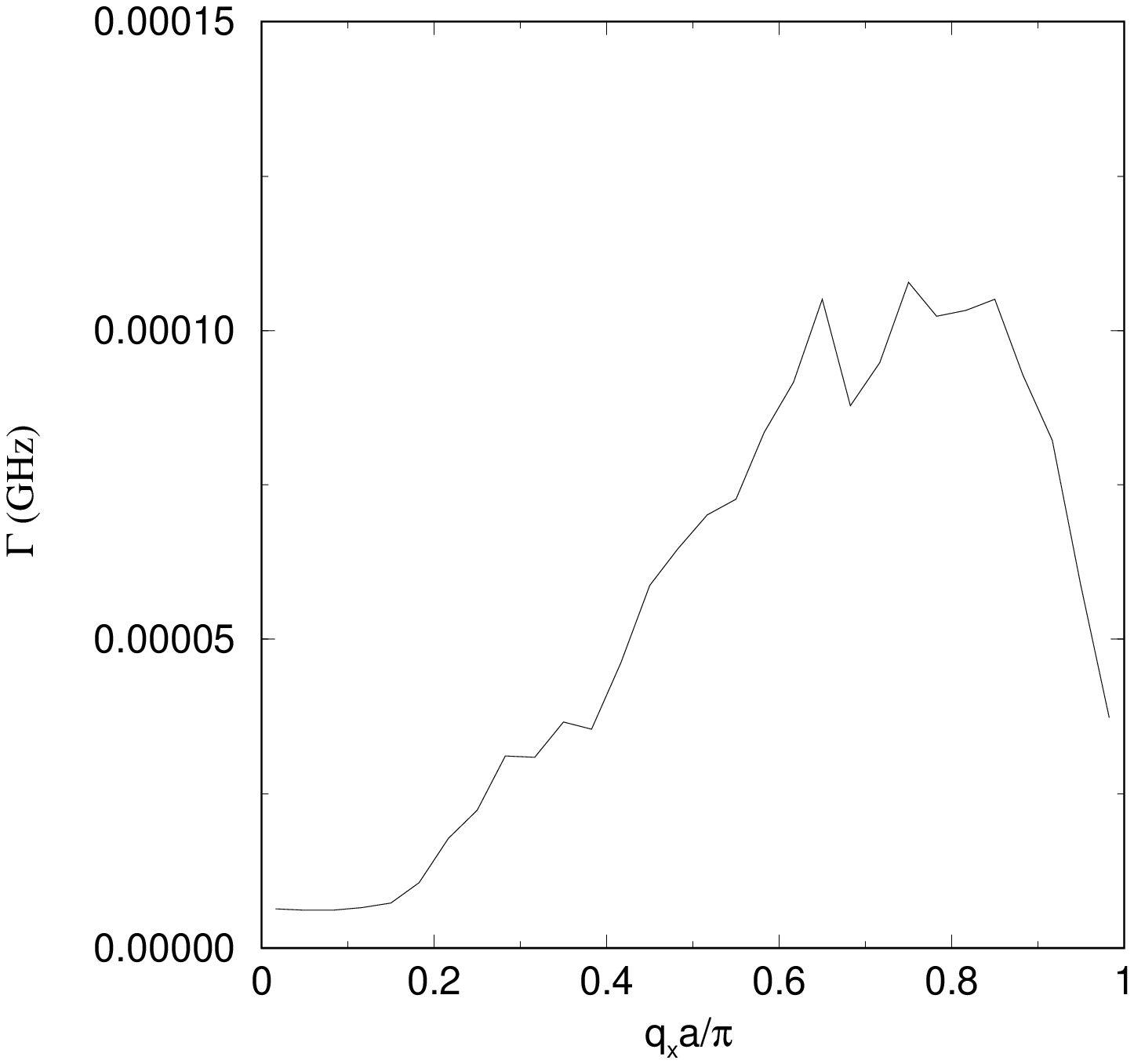}} \vskip5cm
\centerline {Fig. 9.}
\centerline{\epsfysize=12cm\epsfbox{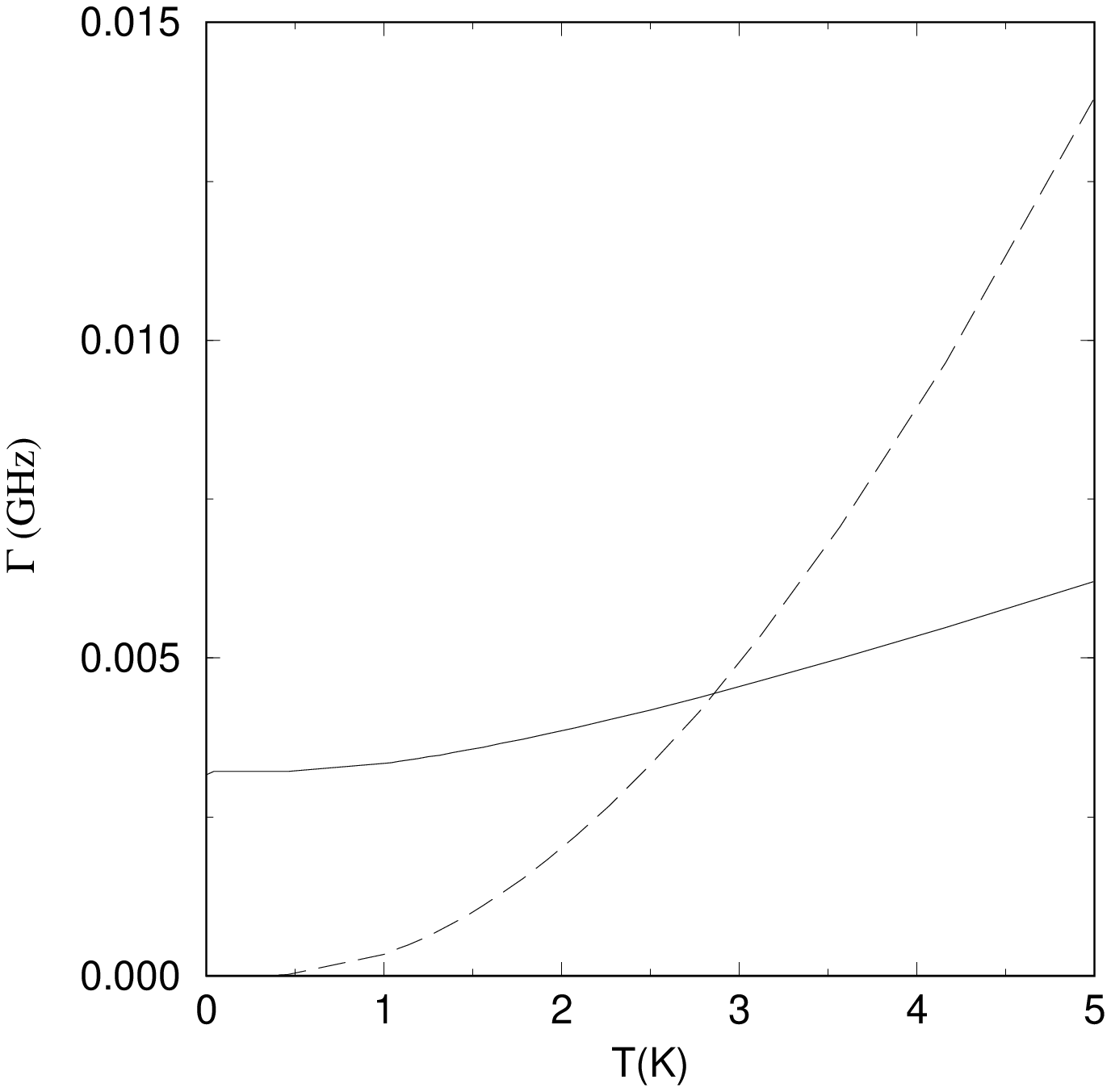}} \vskip5cm
\centerline {Fig. 10.}
\centerline{\epsfysize=12cm\epsfbox{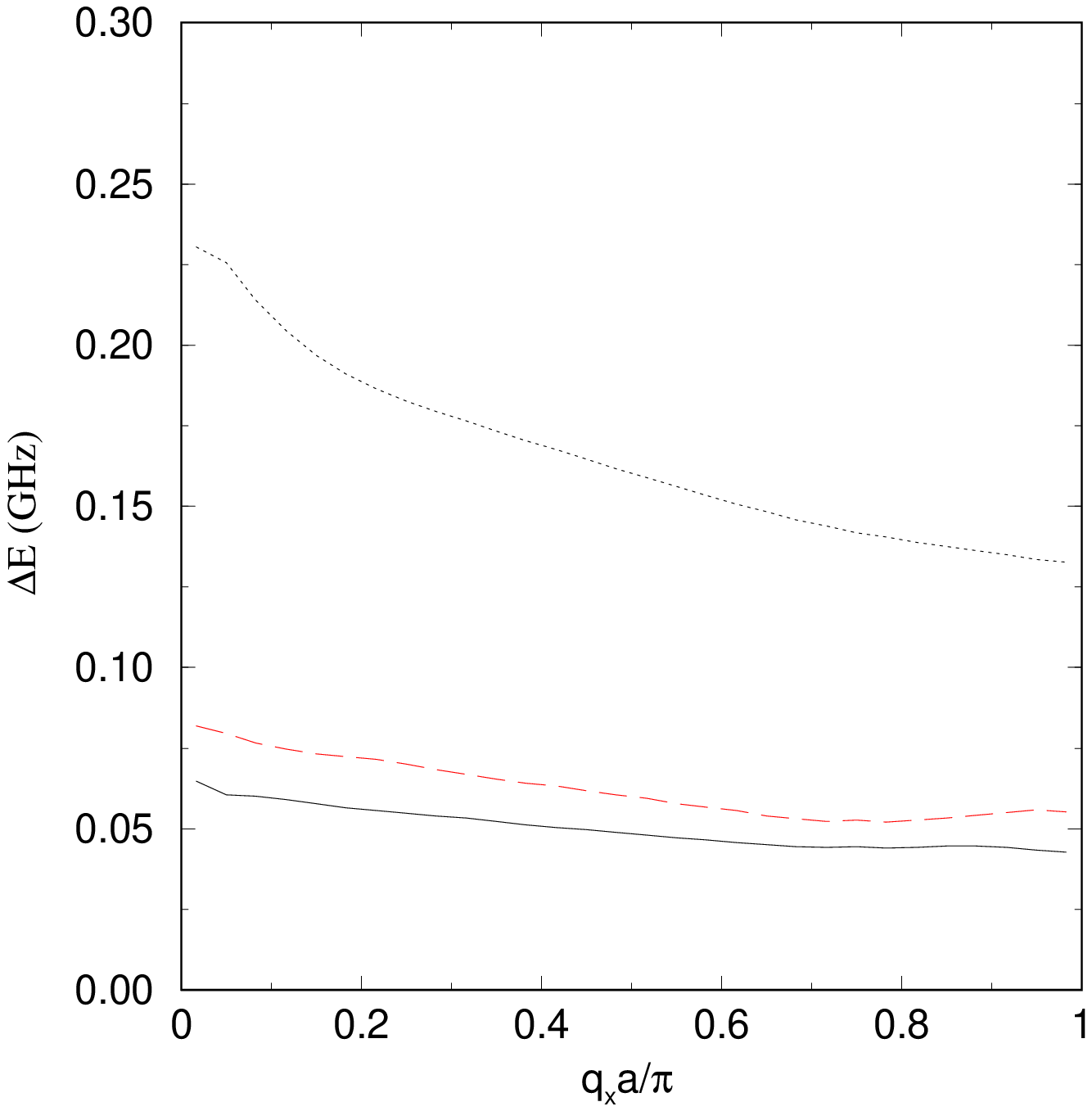}} \vskip5cm
\centerline {Fig. 11.}
\centerline{\epsfysize=12cm\epsfbox{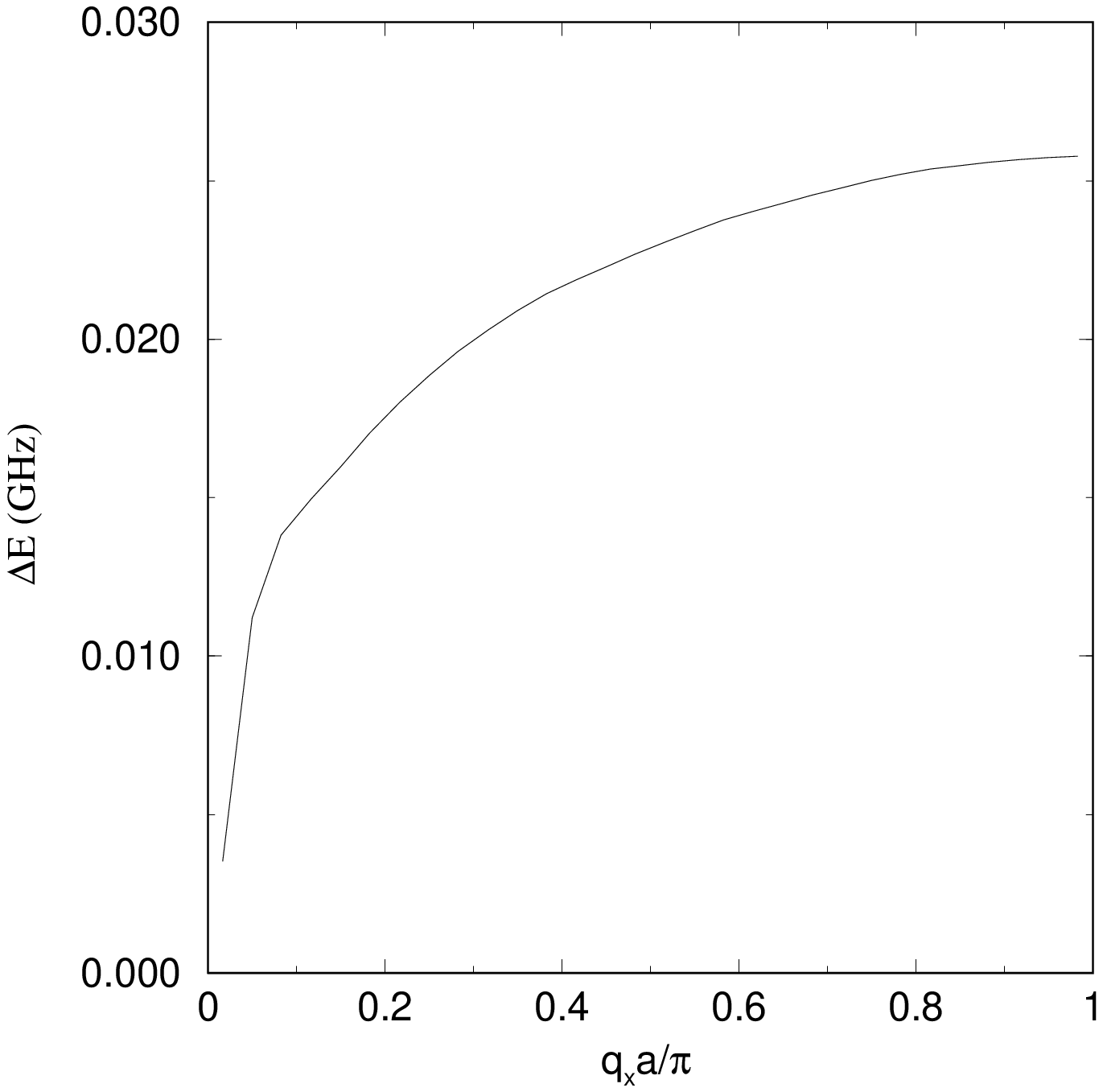}} \vskip5cm
\centerline {Fig. 12.}
\centerline{\epsfysize=12cm\epsfbox{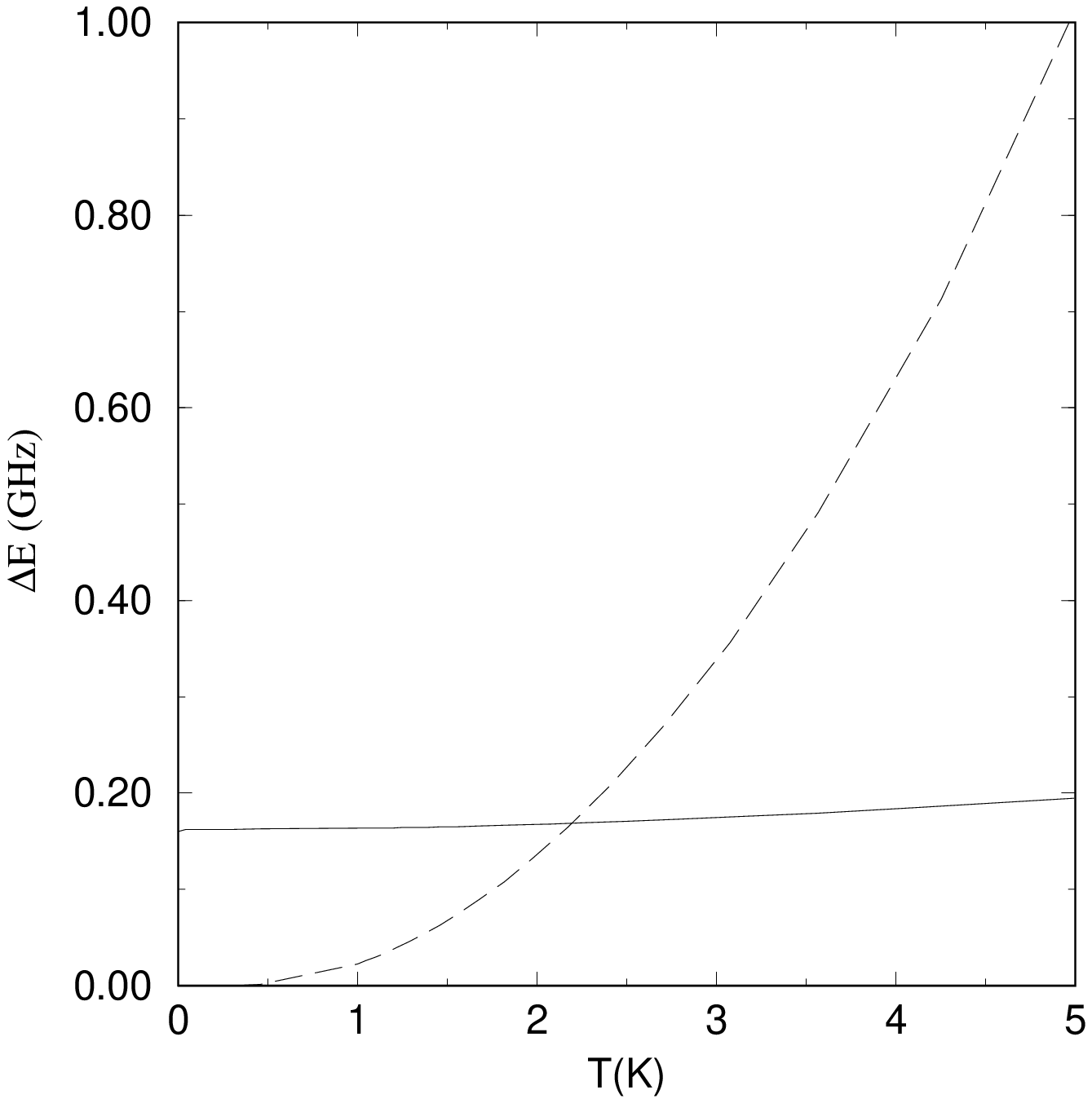}} \vskip5cm
\centerline {Fig. 13.}

\end{document}